\documentclass[twocolumn]{autart}    % Enable this line and disable the 
\usepackage{wrapfig}
\usepackage{graphicx}      % include this line if your document contains figures
\usepackage{natbib}        % required for bibliography

\usepackage{amsmath}
\usepackage{amssymb}
\usepackage{bbm}
\usepackage{xcolor}
\usepackage{circuitikz}
\usetikzlibrary{positioning}
\usetikzlibrary{circuits.ee.IEC}

\newcommand{\tr}{\operatorname{tr}}
\newcommand{\diag}{\operatorname{diag}}
\newcommand{\st}{\, | \,}
\newcommand{\be}{\begin{equation}}
\newcommand{\ee}{\end{equation}}
\newcommand{\R}{\mathbb R}
\newcommand{\C}{\mathbb C}

\DeclareMathOperator{\vol}{volume}

\newtheorem{Theorem}{Theorem}
\newtheorem{Corollary}{Corollary}
\newtheorem{Definition}{Definition}
 \newtheorem{Remark} {Remark}
 \newtheorem{Example} {Example}

\usepackage{setspace}
\usepackage{comment}

\begin{document}
\begin{frontmatter}

\title{Contraction and $k$-contraction in Lurie~systems  with applications to networked systems}

\thanks[footnoteinfo]{This research was partly supported by a research grant from the~Israel Science Foundation.  The work of AO was partly  supported by a research grant from the Ministry of Aliyah and Integration. An abridged version of this paper was accepted for presentation
at the IFAC World Congress 2023~\citep{ron_ifac_version}.} 

\author[First]{Ron Ofir} 
\author[Second]{Alexander Ovseevich}
\author[Second]{Michael Margaliot}

\address[First]{The Andrew and Erna Viterbi Faculty of Electrical and Computer Engineering, Technion---Israel Institute of Technology, Haifa 3200003, Israel (e-mail: rono@campus.technion.ac.il).}
\address[Second]{School of Electrical Engineering, Tel Aviv University, Israel 69978 (e-mails: ovseev@gmail.com, michaelm@tauex.tau.ac.il)}

\begin{abstract}                % %Abstract of not more than 250 words.
%%%%%%%%%%%%%%%%
A  Lurie system  is the interconnection of a linear time-invariant system and a 
nonlinear feedback function.  
We derive a new  sufficient condition for $k$-contraction of a Lurie system. For~$k=1$, our sufficient 
condition reduces to the standard stability condition based on the bounded real lemma and a small gain condition.
 However, Lurie systems often have more than a single equilibrium and are thus not contractive with respect to any norm.
 For~$k=2$, our condition
 guarantees a well-ordered  asymptotic behaviour of the closed-loop system:   every bounded solution converges to an equilibrium, which is not necessarily unique.
 We demonstrate  our results by deriving 
 a sufficient condition for $k$-contraction
 of a general  networked system, and then  applying   it to guarantee  $k$-contraction in a Hopfield neural network,  a nonlinear opinion dynamics model, and a 2-bus power system. 
 %%%%%%%%%%%%%%%
\end{abstract}


\end{frontmatter}
%===============================================================================

\section{Introduction}
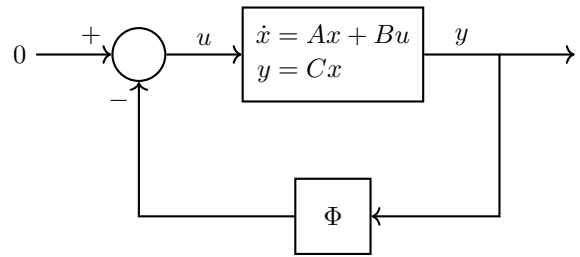
\begin{figure}
    \centering
    \begin{tikzpicture}[
        block/.style = {draw, rectangle, thick, minimum height=2em, minimum width=3em},
        sum/.style = {draw, circle, thick, minimum width=2em}]
        
        % Draw subsystems and sum
        \node[block, minimum height=1.25cm, minimum width=2.4cm] (LTI) {$\begin{aligned}\dot{x} &= Ax + Bu \\ y &= Cx\end{aligned}$} ;
        \node[block, minimum height=1cm, minimum width=1cm] (nonlin) [below=1cm of LTI] {$\Phi$} ;
        \node[sum, left=1cm of LTI] (fbsum) {};
        
        % Draw signals
        \draw[->, thick] (LTI.east) -- node[above, midway] {$y $} +(1,0) coordinate(LTIaux) |- (nonlin.east);
        \draw[->, thick] (LTIaux) -- +(1,0);
        \draw[->, thick] (nonlin.west) -| (fbsum.south) node[below left] {$-$};
        \draw[->, thick] (fbsum.east) -- (LTI.west) node[midway,above] {$u $};
        \draw[<-, thick] (fbsum.west) node[above left] {$+$} -- +(-1,0) node[left] {$0$};
        \end{tikzpicture}
    \caption{Block diagram of a Lurie system.}
    \label{fig:lurie}
\end{figure}

Consider a nonlinear system obtained by connecting
a linear time-invariant~(LTI) system with state vector~$x\in \R^n$, input~$u\in\R^m$ and output~$y\in \R^q$:   
\begin{equation}\label{initial}
\begin{array}{l}
\dot x(t)=Ax(t)+ Bu(t) ,\\%[1em]
y(t)=Cx(t) ,
\end{array}
\end{equation}
%%%%%%%%%%%%%%%%%
with a time-varying 
nonlinear   feedback control 
\[u(t)=-\Phi(t,y(t)) 
\]
(see Fig.~\ref{fig:lurie}). The resulting closed-loop system  
\begin{equation}\label{eq:closed_loop1}
    \dot x(t) = Ax(t) - B\Phi(t,Cx) .
\end{equation}
is known as  a  Lurie (sometimes written Lure, Lur'e or Lurye)  system after the Russian mathematician Anatolii Isakovich Lurie.

The  non-trivial and  well-studied   absolute 
stability problem 
is to prove  that the closed-loop system is asymptotically stable for any $\Phi$ belonging to a certain  class   of nonlinear functions, e.g., the class of sector-bounded functions~\citep[Ch.~7]{khalil_book}.  

In the 1940s and 1950s,   M. Aizerman and   R. Kalman
conjectured that for certain classes of non-linear functions the  absolute stability problem
can be reduced to the stability analysis of certain classes of linear systems.  
%%%%%%
These conjectures are now known to be false. However, the study of the absolute stability problem has led to many important developments including: 
%%
%%%
(1)~sufficient  conditions for absolute stability in terms of 
the transfer function of the linear system and their graphical interpretations~\citep{khalil_book,vidyasagar2002nonlinear}; 
(2)~passivity-based analysis of interconnected systems, and the so-called Zames–Falb multipliers~\citep{CARRASCO20161};
(3)~the theory of integral quadratic constraints~(ICQs)~\citep{ICQ};
and
(4)~the  formulation of an optimal control approach in the stability analysis of switched linear systems~(see the survey paper~\citep{MARGALIOT20062059}).

Several authors studied~\eqref{eq:closed_loop1} using contraction theory. A system is called  contractive  if any two trajectories approach each other at an exponential rate~\citep{LOHMILLER1998683,sontag_cotraction_tutorial}. In particular, if an equilibrium exists then it  is  unique and  globally exponentially asymptotically stable. 
 \cite{Smith1986haus} derived  a sufficient condition for what is now known as $\alpha$-contraction~\citep{wu2020generalization}, with~$\alpha$ real,  with respect to (w.r.t.) Euclidean norms, applied it to a Lurie system, and demonstrated the results by  bounding  the Hausdorff dimension of attractors of the Lorentz equation. However, his sufficient condition is
  highly conservative, especially for large-scale systems. \cite{Andrieu2019LMIContract}  provide a linear matrix inequality (LMI) sufficient condition for contraction w.r.t.
  Euclidean norms under differential sector bound or monotonicity assumptions on the non-linearity (see also~\cite[Theorem 3.24]{bullo_contraction} for a similar condition under different  assumptions), and use it to design controllers which guarantee contraction of  the closed-loop system. \cite{control_syn_vincent} showed  that the designed controllers yield a closed-loop system with the desirable property of 
  infinite gain margin.  \cite{Proskurnikov2022GeneralizedSLemma} provide  a sufficient condition for contraction w.r.t. non-Euclidean norms (see also~\cite{AD-AVP-FB:21k} where this question was studied in the context of recurrent neural networks). However, 
a Lurie system may have more than a single equilibrium point (see, e.g.~\citep{MIRANDAVILLATORO201876} which studies such systems using dominance theory~\citep{Forni2019diff_diss}),
and then it is not contractive w.r.t. any norm.

Following the seminal work   of~\cite{muldowney1990compound},
\cite{kordercont} recently 
introduced  the notion of  $k$-contractive systems. 
Classical  contractivity implies  that under the phase flow of the system  the tangent vectors to the phase space contract exponentially fast; $k$-contactivity implies that  the same property holds for   elements of $k$-exterior  powers of the tangent spaces. Roughly speaking, this is equivalent to the fact that the flow of the  variational  equation  contracts $k$-dimensional parallelotopes  at an exponential rate. In particular, a~$1$-contractive system is just a contractive system. 
However, a system  that is $k$-contractive, with~$k>1$, may not be contractive in the standard sense. For example, every bounded solution of a time-invariant $2$-contractive system  
converges to an equilibrium point, which may not be unique~\citep{li1995}. Thus, 2-contraction may be useful for analyzing multi-stable systems that cannot be analyzed using standard contraction theory.

The basic tools required to define and study~$k$-contractivity are the~$k$-multiplicative and~$k$-additive compounds of a  matrix. 
  The reason for this is simple: 
$k$-multiplicative compounds  provide information on the volume of  parallelotopes generated by~$k$ vertices, and 
$k$-additive compounds  describe the dynamics of $k$-multiplicative compounds, when the vertices follow
a linear dynamics~\citep{comp_long_tutorial}. 

Here, we derive a novel sufficient condition for~$k$-contracti\-vi\-ty of a Lurie system with respect to a weighted Euclidean norm. A unique feature of this condition is that it
combines an algebraic Riccati inequality~(ARI)
that includes $k$-additive compounds of the matrices of the~LTI, and a kind of gain condition on the Jacobian~$J_\Phi$ of the nonlinear function~$\Phi$. We refer to this special ARI as the~$k$-ARI.

In the special case~$k=1$, the~$k$-ARI reduces to the standard Hamilton-Jacobi 
inequality appearing in the small gain theorem~\cite[Ch.~5]{khalil_book}, and our contraction condition     reduces to a small-gain sufficient condition for standard contraction. However, for~$k>1$ 
our condition provides new results. 
We demonstrate  this  by deriving a simple sufficient  condition for~$k$-contraction of a general networked  system and then applying it to a Hopfield neural network, a nonlinear opinion dynamics model, and a 2-bus power system. These  systems  are typically multi-stable,
and thus
cannot be analyzed using standard contraction theory. Nevertheless, for the case~$k=2$ our sufficient condition still guarantees a well-ordered global
behaviour: any bounded solution converges to an equilibrium point, that is not necessarily unique.

We use standard notation. For a square matrix~$A\in\C^{n\times n}$, $\tr (A)$ is the trace of~$A$, and~$\det (A)$ is the determinant 
of~$A$. 
$A^*$ is the   conjugate
transpose of~$A$. If~$A$ is real then
this is just the transpose of~$A$, denoted
$A^T$.  
A symmetric matrix~$P \in \R^{n \times n}$ is called positive definite [positive semi-definite] if $x^TPx > 0$ [$x^TPx \ge 0$] for all~$x \in \R^n\setminus\{0\}$. Such matrices are denoted by $P \succ 0$ and $P \succeq 0$, respectively. For~$A \in  \C^{n \times m}$, $\sigma_1(A) \ge \dots \ge \sigma_{\min\{n,m\}}(A) \ge 0$   
denote the ordered singular values of~$A$, that is, the ordered 
square roots of the eigenvalues of~$A^*A$ if $m<n$, or of~$AA^*$, otherwise. The~$n\times n$ identity matrix is denoted by~$I_n$. 
The~$L_2$  norm  of a vector~$x$ is~$|x|_2:=(x^Tx)^{1/2}$, and the induced~$L_2$ norm of a matrix~$A$ is~$\|A\|_2 = \sigma_1(A)$. For two integers
$i\leq j$, we let~$[i,j]:=\{i,i+1,\dots,j\}$.

The remainder of this paper is organized as follows. The next section reviews two basic tools used to establish   $k$-contraction: matrix compounds and matrix measures. Section~\ref{sec:main}
presents and discusses the  main result. 
Section~\ref{sec:proof} proves the main result.
Section~\ref{sec:networked} describes an application of our main result to a networked system and demonstrates
how this can be used to analyze $k$-contraction in
a Hopfield neural network, a nonlinear opinion dynamics model, and a 2-bus power system. 
The final section concludes.

\section{Preliminaries}
%%%%%%%%%%%%%%%%%%%
In this section, we review several known definitions and results on  matrix  compounds and matrix measures that will be used in Section~\ref{sec:main}.
%%%%%%%%%%%%%%%%%%%
 \subsection{Matrix compounds}
%%%%%%%%%%%%%%%%%%%%%%%%%%%%%%%
  For two integers~$i,j$, with~$ i\leq j$, let~$[i,j]:=\{i,i+1,\dots,j\}$. 
 Let~$Q_{k,n}$ denote the set 
of increasing sequences of~$k$ numbers from~$[1,n]$
ordered lexicographically.
For example,~$Q_{2,3} =
\{ (1,2), (1,3), (2,3) \}
$. 
%%%%%%%
%%

For~$A\in\R^{n\times m}$ and~$k\in[1,\min\{n,m\}]$, 
  a \emph{minor of order~$k$} of~$A$ is the determinant of some~$k \times k$
submatrix of~$A$.  
%%%%%%%
%%%%%%%%%
%%%%%%%%%
Consider the~$\binom{n}{k}\times \binom{m}{k}  $
 minors of  order~$k$ of~$A$. 
Each such minor is defined by a set of row indices~$\kappa^i \in Q_{k,n}$ and column indices~$\kappa^j\in 
Q_{k,m}$. This minor 
is denoted by~$A(\kappa^i|\kappa^j)$.
%%%%%%
For example, for~$A=\begin{bmatrix} 1&2   \\ -1 &3   \\0&3 
\end{bmatrix}$,  we have
$
A((1,3) |(1,2))=\det  \begin{bmatrix} 1&2\\0&3
\end{bmatrix}   =3.
$
%%%%%%%%%%%
\begin{Definition}\label{def:multi} 
    The~$k$-\emph{multiplicative compound matrix} 
of~$A\in\R^{n\times m}$, denoted~$A^{(k)}$, is the~$\binom{n}{k}\times  \binom{m}{k}$ matrix
that includes all  the minors of order~$k$ ordered lexicographically.
%%%%%%%
\end{Definition} 
%%%%%%
For example, for~$n = m =3$ and~$k=2$, we have 
\[
%%%
	 A^{(2)}= \begin{bmatrix}
                   A((1,2)|(1,2)) & A((1,2)|(1,3)) & A((1,2)|(2,3))\\
						A((1,3)|(1,2)) & A((1,3)|(1,3)) & A((1,3)|(2,3))\\
						A((2,3)|(1,2)) & A((2,3)|(1,3)) & A((2,3)|(2,3)) 
\end{bmatrix}.
%%%
\] 
%%%%%%%%%%%%%%%%%%%%%%%%%%%%%%%%%%%%%%%%%%%%%%
Definition~\ref{def:multi} has several implications. First, if~$A$ is square then~$(A^T)^{(k)} = (A^{(k)})^T$, and in particular if~$A$ is symmetric then so is~$A^{(k)}$. 
Also, $A^{(1)}=A$ 
and if~$A \in \mathbb{R}^{n \times n}$ then~$A^{(n)}=\det(A)$.
If~$D$ is an~$n\times n$ diagonal matrix, i.e.~$D=\diag(d_1,\dots,d_n)$ then
$
D^{(k)}=\diag(d_1\dots d_k, d_1\dots  d_{k-1}d_{k+1},\dots,d_{n-k+1}\dots d_n)
$.
In particular, every eigenvalue of~$D^{(k)}$ is the product of~$k$ eigenvalues of~$D$. 
In the special case~$D= p I_n$, with~$p \in \R$, we have that~$(pI_n)^{(k)}=p^k I_r$, with~$r:=\binom{n}{k}$.  
 
The \emph{Cauchy-Binet formula}  (see, e.g.,~\cite[Thm.~1.1.1]{total_book}) asserts   
that 
\be\label{eq:cbf}
%%%
(AB)^{(k)}=A^{(k)} B^{(k)}  
\ee
  for any $A \in \R^{n \times p}$, $B \in \R^{p \times m}$, $k \in [1,\min\{n,p,m\} ]$.
  This justifies the term \emph{multiplicative compound}.  
  
When~$n=p=m=k $, Eq.~\eqref{eq:cbf}  becomes the familiar formula~$\det(AB)=\det(A)\det(B)$.
If~$A$ is $n\times n $ and non-singular then~\eqref{eq:cbf}  implies that $ I_n^{(k)}=(AA^{-1})^{(k)}=A^{(k)} (A^{-1})^{(k)}$,
	so~$A^{(k)} $ is also non-singular  with
 \[
 (A^{(k)})^{-1}=(A^{-1})^{(k)}   .
 \]
	Another implication of~\eqref{eq:cbf}
	is that if~$A\in\R^{n\times n}$ with eigenvalues~$\lambda_1,\dots,\lambda_n$
then the eigenvalues of~$A^{(k)}$ are all the~$\binom{n}{k}$ products:
\[
\lambda_{i_1} \lambda_{i_2} \dots \lambda_{i_k}, \text{ with } 1\leq i_1 < i_2 <\dots < i_k\leq n.
\]
	
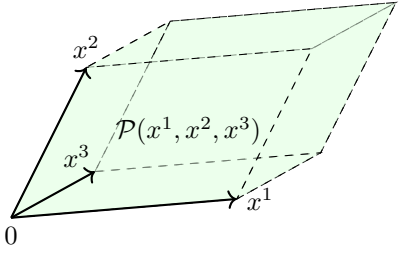
\begin{figure}
	    \centering
	    \begin{tikzpicture}
            \draw[dashed,fill opacity=0.5,fill=green!10] (0,0,0)--(3,0.25,0)--(4.5,1.25,1)--(1.5,1,1)--cycle;
            \draw[dashed,fill opacity=0.5,fill=green!10] (1,2,0)--(4,2.25,0)--(5.5,3.25,1)--(2.5,3,1)--cycle;
            \draw[dashed,fill opacity=0.5,fill=green!10] (3,0.25,0)--(4.5,1.25,1)--(5.5,3.25,1)--(4,2.25,0)--cycle;
            \draw[dashed,fill opacity=0.5,fill=green!10] (0,0,0)--(1.5,1,1)--(2.5,3,1)--(1,2,0)--cycle;
            \draw[dashed,fill opacity=0.5,fill=green!10] (1.5,1,1)--(4.5,1.25,1)--(5.5,3.25,1)--(2.5,3,1)--cycle;
            \draw[dashed,fill opacity=0.5,fill=green!10] (0,0,0)--(3,0.25,0)--(4,2.25,0)--(1,2,0)--cycle;
            
            \draw[thick,->] (0,0,0)--(3,0.25,0) node[right]{$x^1$};
            \draw[thick,->] (0,0,0)--(1,2,0) node[above]{$x^2$};
            \draw[thick,->] (0,0,0)--(1.5,1,1) node[above left=-0.1cm]{$x^3$};
            \draw (0,0,0) node[below]{0};
            \draw (2.55,1.35,0.5) node[]{$\mathcal{P}(x^1,x^2,x^3)$};
        \end{tikzpicture}
	    \caption{A 3D parallelotope with vertices $0,x^1,x^2$, and $x^3$.}
	    \label{fig:parallelogram}
	\end{figure}

	The usefulness of the $k$-multiplicative compound in analyzing 
	$k$-contraction follows from the   relation between the $k$-compound and the volume of~$k$-parallelotopes. To explain this, 
	fix~$k$ vectors~$x^1,\dots,x^k \in \R^n$. The parallelotope  generated by these vectors (and the zero vertex) is
	\[
	\mathcal{P}(x^1 , \dots,x^k) := \left\{\sum_{i=1}^k r_i x^i \st r_i \in [0,1] \text { for all } i \right\},
	\]
	(see   Fig.~\ref{fig:parallelogram}).
	Let
	\[
	X:=\begin{bmatrix} x^1&\dots&x^k
	\end{bmatrix} \in \R^{n\times k}.
	\]
	The volume of~$\mathcal{P}(x^1,\dots,x^k)$ satisfies~\citep[Chapter~IX]{Gantmacher_vol1}:
	\be\label{eq:voldet}
	\vol (\mathcal{P}(x^1,\dots,x^k))= |X^{(k)} |_2.
	\ee
	Note that since~$X\in\R^{n\times k}$, the dimensions of~$X^{(k)}$ are~$\binom{n}{k}\times 1$, that is, $X^{(k)}$ is a column vector. 
	
	\begin{Example}
	    Consider the case~$n=3$, $k=2$, $x^1=\begin{bmatrix} a& 0& 0 
	    \end{bmatrix}^T$, and~$x^2=\begin{bmatrix} 0& b& 0 
	    \end{bmatrix}^T$, with~$a,b\in \R$. Then
	    $X=\begin{bmatrix}
	   a&0 \\0&b \\0&0 
	    \end{bmatrix}
	    $, so
	    $X^{(2)}=\begin{bmatrix}
	    ab&0&0 
	    \end{bmatrix}^T
	    $,  and~$|X^{(2)}|_2=|ab|$.
	\end{Example}

	In the special case~$k=n$, Eq.~\eqref{eq:voldet}
	becomes the well-known formula 
	\begin{align*}
	    \vol (\mathcal{P}(x^1,\dots,x^n))&= |X^{(n)}  |_2\\&=|\det(X)|.
%%%%%%	
	\end{align*}

	When the   vertices of the parallelotope
	follow  a linear time-varying dynamics, the evolution of the $k$-multiplicative compound depends on another algebraic construction called the $k$-additive compound. 
	%%%%%%%%%%%%%%%%%%%%%%%%%%%%%%%%%%%%
 %%%%%%%%%%%%%%%%%%%%%%%%%%%%%%%%%
\begin{Definition}\label{def:addi_comp}
    The $k$-\emph{additive compound matrix} of~$A \in \R^{n \times n}$
is   defined  by
\begin{align} 
    			A^{[k]} := \frac{d}{d \varepsilon}  (I_n+\varepsilon A)^{(k)} |_{\varepsilon=0}  .
\end{align}
\end{Definition}
Note that this implies that~$	A^{[k]}   =
    			\frac{d}{d \varepsilon}  
    			(\exp(\varepsilon A))^{(k)}|_{\varepsilon=0}$. 
%%%%%%%%%%%%%%%%%%%%%
%%%%%%%%%%%%%%%%
\begin{Example}
    Suppose that~$A=pI_n$, with~$p\in\R$. Then 
   \begin{align*}
  ( I_n+\varepsilon A)^{(k)}  &=( (1+\varepsilon p  ) I_n)^{(k)} \\
  &=(1+\varepsilon p  )^k I_r, 
   \end{align*}
where~$r:=\binom{n}{k}$, so   
  \begin{align*}   
    	(p I_n)^{[k]} &= \frac{d}{d \varepsilon}  (1+\varepsilon p  )^k I_r |_{\varepsilon=0}\\
    	&=k p I_r.
    	\end{align*} 
\end{Example}

Definition~\ref{def:addi_comp} implies that~$A^{[1]}=A$, 
$A^{[n]}=\tr(A)$,
and that 
\be\label{eq:poyrt}
(I_n+\varepsilon A)^{(k)}= I_r+\varepsilon   A^{[k]} +o(\varepsilon )  ,
\ee 
%%%%%%
%%%
where~$r:=\binom{n}{k}$.
Thus,~$\varepsilon  A^{[k]}$ is the first-order term in the Taylor series of~$(I+\varepsilon A)^{(k)}$. 
Also,   $(A^T)^{ [k]} = (A^{[k]})^T$, and in particular if~$A$ is symmetric then so is~$A^{[k]}$. 

\begin{Example}\label{exa:sdig}
If~$D=\diag(d_1,\dots,d_n)$ then
$
(I+\varepsilon D)^{(k)}=\diag \left ( \prod_{i=1}^k (1+\varepsilon d_i) 
 ,\dots,\prod_{i=n-k+1}^n (1+\varepsilon d_i)    \right  ),
$
  so~\eqref{eq:poyrt} gives 
$
D^{[k]}=\diag( \sum_{i=1}^k d_i 
 ,\dots,  \sum_{i=n-k+1}^n d_i
 )     .
$
In particular, every eigenvalue of~$D^{[k]}$ is the sum of~$k$ eigenvalues of~$D$. 
\end{Example}
%%%%%%%%%%%%
%%%%%%%%%%%%%%
 
 More generally,   if~$A\in\R^{n\times n}$ with eigenvalues~$\lambda_1,\dots,\lambda_n$
then the eigenvalues of~$A^{ [ k ] }$ are all the~$\binom{n}{k}$ sums:
\[
\lambda_{i_1} + \lambda_{i_2} + \dots +  \lambda_{i_k} , \text{ with } 1\leq i_1 < i_2 <\dots < i_k\leq n,
\]
(see e.g. \cite[Thm. 6.24]{fiedler_book} or \cite{comp_long_tutorial}).

It follows from~\eqref{eq:poyrt} and the properties of the multiplicative compound that~$
(A+B)^{[k]}= A^{[k]}+B^{[k]} 
$ for any~$A,B\in\R^{n\times n}$,  
thus justifying the term \emph{additive compound}. In fact, the mapping~$A\to A^{[k] }$ is linear~\citep{schwarz1970}.

Note that if~$Q\in\R^{n\times n}$ is positive definite then it is symmetric with positive eigenvalues and thus~$Q^{(k)}$ and~$Q^{[k]}$ are symmetric with positive eigenvalues, so  they are also positive definite.

Below we  will use the following  relations. Let~$A \in \R^{n \times n}$. If $U \in \R^{p \times n}$ and~$V\in \R^{n \times p}$ then
\begin{equation}\label{eq:k_mul_coor_trans}
(U A V  )^{ (k) } = 	U^{(k)} A^{ (k) } V^{ (k) }   , 
\end{equation}
and if, in addition, $UV=I_p$  then combining this with Definition~\ref{def:addi_comp} gives
\begin{equation}\label{eq:k_add_coor_trans}
(U A V )^{[k]} = 	U^{(k)}A^{[k]} V ^{(k)}    .
\end{equation}

  For more on the applications of compound matrices  to
systems and control theory, see e.g.~\citep{wu2021diagonal, margaliot2019revisiting, ofir2021sufficient, ron_DAE,grussler2022variation,LI1999191}, and the recent tutorial by~\cite{comp_long_tutorial}.

%%%%%%%%%%%%%%%%%%%%%%%
\subsection{Matrix measures}
%%%%%%%%%%%%%%%%%%%%%%%%%%%%%%%%%%
Matrix measures (also called logarithmic norms~\citep{strom1975logarithmic}) provide an easy to check  sufficient condition for contraction~\citep{sontag_cotraction_tutorial}. 
%%%
Fix  a   norm~$|\cdot|:\R^n\to\R_+$.
The  induced matrix norm~$\|\cdot\|:\R^{n\times n}\to\R_+$ is 
defined by~$\|A\| := \max_{|x|=1} |Ax| $, and the
 induced   matrix measure   $\mu(\cdot):\R^{n\times n}\to\R $   is
defined by
\[
\mu(A) := \lim_{\varepsilon \downarrow 0} \frac{\|I + \varepsilon A\| - 1}{\varepsilon} .
\]
%%%%
The matrix measure is sub-additive, i.e.
\[
\mu(A+B)\leq\mu(A)+\mu(B).
\]
Also,~$\mu(c I_n)=c$ for any~$c\in \R$. 

The matrix measure induced by the~$L_2$ norm is~\citep{vidyasagar2002nonlinear}:
\begin{equation}\label{eq:matirxm} \begin{aligned} 
\mu_2(A) = (1/2) \lambda_{\max}\left(  A + A^T \right) ,   
\end{aligned} \end{equation}
where~$\lambda_{\max} (S)$ denotes the largest eigenvalue of the symmetric matrix~$S$.

For an invertible matrix~$H \in\R^{n\times n}$, a  scaled~$L_2$ norm is defined by~$|x|_{2,H} :=|H x|_2$, and the induced matrix measure is 
\begin{align}\label{eq:weight_mat_meas}
    \mu_{2,H}(A)& =   
    \mu_2(H A H^{-1} )\nonumber\\& 
    =    (1/2)   \lambda_{\max}\left(H A  H^{-1}  + 
    (H A H^{-1} )^T  \right).
    %%%%%%%%%
\end{align}

Roughly speaking, a system is $k$-contractive  if the volume of $k$-dimensional bodies decays at an exponential rate under the flow of the dynamics. An exact definition may be found in~\cite{kordercont}. For this paper, it is only required to know the following sufficient condition: The system~$\dot x=f(t,x)$ is $k$-contractive if~$\mu((J(t,x))^{[k]})\leq-\eta<0$ for all~$t,x$, where~$J:=\frac{\partial}{\partial x}f$ is the Jacobian of the vector field~$f$. For~$k=1$, this reduces to the standard sufficient condition\footnote{For the case of 1-contraction, this condition is known to be necessary and sufficient under certain  assumptions on the vector field~$f$. However, no such result is currently known for $k$-contraction.} for contraction, namely,  $\mu(J (t,x)) \leq -\eta<0$ for all~$t,x$.
Indeed,~$1$-contraction is just contraction.

Note that if~$A,H\in\R^{n\times n}$, with~$H$ non-singular, then
\begin{align}\label{eq:mu2add}
\mu_{2,H^{(k)}  } (A^{[k]} ) & = \mu_{2  } (H^{(k)}A^{[k]} (H^{(k)}) ^{-1} )\nonumber\\
&=\mu_2(  (HAH^{-1})^ { [k]}    ), 
\end{align}
where the last equality follows from~\eqref{eq:k_add_coor_trans}.

\begin{Example}
    Consider the LTI
\be\label{eq:lti}
\dot x(t)=Ax(t).
\ee
If~$\mu(A ^{[1]} ) <0$ for some matrix measure~$\mu$ then~$A^{[1]}=A$ is Hurwitz, and thus every solution of~\eqref{eq:lti} converges to the unique eqilbrium at the origin. 
If~$\mu(A^{[2]}) <0$ for some matrix measure~$\mu$ then~$A^{[2]}$ is Hurwitz. 
Thus, the sum of any two eigenvalues of~$A$ has a negative real part. In particular,~$A $ cannot have any purely imaginary eigenvalues, so 
any bounded solution of~\eqref{eq:lti} converges to the origin. 
\end{Example} 

\section{Main result}\label{sec:main}
%%%%%%%%%%%%%%
In this section, we derive a sufficient condition for $k$-contraction of the closed-loop system~\eqref{eq:closed_loop1}.
We assume that~$\Phi$ is continuously differentiable and denote its Jacobian
by~$J_\Phi(t,y) := \frac{\partial \Phi}{\partial y}(t,y)$. The Jacobian of~\eqref{eq:closed_loop1} is then
%%%%%%%%%%%%%%%%%%%%%%%%%%%
\begin{equation}\label{eq:closed_loop_jac}
    J(t,x): = A - BJ_\Phi(t,Cx)C, 
\end{equation}
so
\[
J^{[k]}(t,x) = A^{[k]}-(B J_\Phi(t,Cx) C)^{[k]}.
\]
Guaranteeing that~$\mu(J^{[k]}(t,x))\leq-\eta<0$ is non-trivial due to the term~$(BJ_\Phi(t,Cx) C)^{[k]}$. Our goal is to find a sufficient condition  guaranteeing that there exists a weight matrix $P$ such that~$\mu_{2,P^{(k)}}(J^{[k]}(t,x))\leq-\eta<0$ where the condition satisfies the following properties: (1) it decomposes, as much as possible,  
 to a condition on the linear subsystem and a condition on  the non-linearity~$\Phi$; (2) it reduces for~$k=1$ to a standard  sufficient condition for contraction; and (3) for $k>1$ it is strictly weaker than the 
standard  sufficient condition for contraction, that is,  $\mu(J(t,x))\leq -\eta<0$.

We can now state our main result. For a symmetric matrix~$S\in\R^{n\times n}$,  we denote its ordered eigenvalues by~$\lambda_1(S)\geq\dots\geq \lambda_n(S)$.
%%%%%%%%%%%%%%%%%%%%%
\begin{Theorem}\label{thm:lure_suff}
%%%%%%%%%%%%%%%%%%%%%%%%%%%%%%%%%%%%%%%%
Consider the Lurie system~\eqref{eq:closed_loop1}.
	Fix~$k \in [1,n]$.
	Suppose  that
	there exist~$\eta_1,\eta_2\in \R$ and~$P \in \R^{n \times n}$, where~$P =QQ $ with~$Q \succ 0$, such that
 %%%%
	\begin{align}\label{eq:k_riccati_Pk}
    &    P^{(k)}A^{[k]}  + (A^{[k]})^TP^{(k)} + \eta_1 P^{(k)} \\
        &+ Q^{(k)}\left((QBB^TQ)^{[k]} + (Q^{-1}C^TCQ^{-1})^{[k]}\right)Q^{(k)} \preceq 0 , \nonumber
    \end{align}
and, furthermore, at least one of the   following  two conditions hold:
%%%%%%
	\begin{equation}\label{cond:k_smallgain_CQ}
		\sum_{i=1}^k \lambda_i\left( Q^{-1}C^T \left (
  (J_\Phi^T(t,y)  J_\Phi(t,y) - I_q
\right   )  CQ^{-1} \right ) \leq - \eta_2 ,
	\end{equation}
 %%%%
 or
 \begin{equation}\label{cond:k_smallgain_BQ}
		\sum_{i=1}^k \lambda_i\left( QB  \left (
  (J_\Phi(t,y)  J_\Phi^T(t,y) - I_m
\right   )  B^TQ \right ) \leq - \eta_2 ,
	\end{equation}
%%%%    %%%%%    
	for all $t \ge 0, y \in \R^q$.
 Then the Jacobian of the closed-loop system~\eqref{eq:closed_loop1} satisfies
	\[
	\mu_{2,Q^{(k)}}(J^{[k]}(t,x)) \leq  -(\eta_1 + \eta_2)/2 \text{ for all } t \ge 0, x \in \R^n.
	\]
  In particular, if~$\eta_1 + \eta_2 > 0$, then the closed-loop system~\eqref{eq:closed_loop1} is $k$-contractive with rate~$(\eta_1 + \eta_2)/2$ w.r.t. the scaled $L_2$ norm~$|z|_{2,Q^{(k)}}=|Q^{(k)}z|_2$.
\end{Theorem}

Before proving this result (see  Section~\ref{sec:proof}), we give  several comments.

We refer to condition~\eqref{eq:k_riccati_Pk} as the~$k$-ARI. Note that this condition only involves the matrices $A,B,C$ defining the linear subsystem.
Conditions~\eqref{cond:k_smallgain_CQ} and~\eqref{cond:k_smallgain_BQ} include both the matrices~$B,C,Q$ and the Jacobian of the non-linear function. However, if
the small gain condition~$\sigma_1(J_\Phi)\leq 1$ holds 
then~\eqref{cond:k_smallgain_CQ} and~\eqref{cond:k_smallgain_BQ} both hold with~$\eta_2=0$. More generally, if~$\sigma_1(J_\Phi)$ is uniformly
bounded by some bound~$q$ 
then we can always scale the closed-loop system~\eqref{eq:closed_loop1}
so 
that the small gain condition holds by considering 
\begin{equation}\label{initial_q}
\begin{array}{l}
\dot x=Ax+ qBu ,\\%[1em]
y=Cx,\\
u=-\frac1{q}\Phi(t,y).
\end{array}
\end{equation}
Now applying Thm.~\ref{thm:lure_suff} yields the following result.

\begin{Corollary}
%%%%%%%%%%%%%%%%%%%%%%%%%%%%%%%%%%%%%
Suppose that 
\be\label{eq:condj}
\sigma_1(J_\Phi(t,y))\leq q \text{ for all } t\geq 0 
\text{ and } y \in \R^q,
\ee
and   that
	there exist~$\eta_1 >0 $ and~$P \in \R^{n \times n}$, where~$P =QQ $, with~$Q \succ 0$, such that
\begin{align}     \label{eq:k_riccati_Pk_2}
 &    P^{(k)}A^{[k]}   + (A^{[k]})^TP^{(k)} + \eta_1 P^{(k)}\nonumber \\
     &
        + Q^{(k)}\left(q^{2k}(QBB^TQ)^{[k]} + (Q^{-1}C^TCQ^{-1})^{[k]}\right)Q^{(k)} \preceq 0 . 
\end{align}
Then the closed-loop system~\eqref{eq:closed_loop1} is $k$-contractive with rate~$\eta_1/2$ w.r.t. the scaled $L_2$ norm~$|z|_{2,Q^{(k)}}=|Q^{(k)}z|_2$.
%%%%%%%%%%%%%%%%%%%%%%%%%%%%
\end{Corollary}

Note that now the conditions are decoupled: condition~\eqref{eq:condj} refers to the nonlinear feedback,  whereas~\eqref{eq:k_riccati_Pk_2} is a condition on the LTI system.

 \begin{Remark}
%%%%%%%%%%%%
Note that when~$k=1$, Eq.~\eqref{eq:k_riccati_Pk} holds for some~$\eta_1 > 0$ if and only if the familiar ARI
\be\label{eq:fari}
 P A  + A^TP  + P BB^T P  +  C^TC  \prec 0
\ee
holds. Assuming that  the LTI subsystem is minimal,~\eqref{eq:fari} holds if and only if $A$ is Hurwitz and the $H_\infty$ norm of the~LTI subsystem is smaller  than one~\cite[Chapter~5]{khalil_book}. Similarly,~\eqref{cond:k_smallgain_CQ} and~\eqref{cond:k_smallgain_BQ} hold for any $\eta_2 > 0$ if and only if~$\|J_\Phi\|_2 \le 1$, so in the special case~$k=1$ Thm.~\ref{thm:lure_suff} becomes a small-gain sufficient condition for standard contraction.
%%%%%%%%%%%%
\end{Remark}

\begin{Remark}
%%%%%%%%%%%%%%%%%%
Denote
%%%%%%%%%%%%
\begin{align}\label{eq:defmatr}
S &:= QAQ^{-1} +Q^{-1}A^TQ   + {\eta_1}{k^{-1}} I_n +  QB B^TQ\nonumber \\&+  Q^{-1}C^TCQ^{-1} .
\end{align}
Then
\begin{align*}
    S^{[k]}&= Q^{(k)}A^{[k]}(Q^{(k)})^{-1}  + (Q^{(k)})^{-1}(A^{[k]})^TQ^{(k)} + \eta_1 I_r\\& + (QBB^TQ)^{[k]} + (Q^{-1}C^TCQ^{-1})^{[k]} ,
%%%%
\end{align*}
and this implies that 
 condition~\eqref{eq:k_riccati_Pk} can be written more succinctly as
\be\label{eq:rkk}
S^{[k]} \preceq 0 , 
\ee
that is, $\sum_{i=1}^k\lambda_i (S) \leq 0$.
%%%%%%%%%%%%%%%%%%%%%
%%%%%%%%%%%%%%%%%
Consider the particular choice~$P=p I_n$, with~$p>0$. Then~$Q=p^{1/2} I_n$, so
\[
S=A+A^T+\eta_1k^{-1}I_n +p B B^T +p^{-1} C^T C ,
\]
and~\eqref{eq:rkk} becomes
\be\label{eq:almostricca}
A^{[k]}+(A^{[k]})^T +\eta_1 I_r+p (BB^T)^{[k]} +p^{-1}(C^TC)^{[k]}\preceq 0.
 \ee
 Intuitively speaking, this requires~$A^{[k]}+(A^{[k]})^T$ to be  negative-definite ``enough'', so that it remains negative semi-definite even after adding 
  positive semi-definite
 terms related to the input and output channel. 
%%%%%%%%%%%%%%5
\end{Remark}
%%%%%%%%%%%%%%%%

 %%%%%%%%%%%%%%%%%%%
It is natural to expect that a sufficient condition  for~$k$-contraction implies~$\ell$-contraction for any~$\ell > k$ (see  \citep{kordercont,wu2020generalization}).  The next result shows that this is indeed so for the conditions in Theorem~\ref{thm:lure_suff}. 
\begin{prop}
    Suppose that the conditions in Theorem~\ref{thm:lure_suff} hold for some integer~$k\geq 1$  and~$\eta_1,\eta_2\geq 0$. 
    Then  they hold for any~$\ell>k $ with the same~$\eta_1,\eta_2$. 
\end{prop}
%%%%%%%%%%%%%%%%%%%%%%%%%%%%%

\begin{pf}
%%%%%%%%%%%%%%%%%%%%%%%%%%%%%%%%%%%%%%%%%%%%%%%%%%%%%%%
Suppose that   there exists~$P=QQ$, with~$Q\succ 0$, such that~\eqref{eq:k_riccati_Pk} holds with~$\eta_1 \ge 0$, and either~\eqref{cond:k_smallgain_CQ} or~\eqref{cond:k_smallgain_BQ} hold with~$\eta_2\geq  0$. Fix an integer~$\ell>k$.
     Recall that 
     condition~\eqref{eq:k_riccati_Pk} is equivalent to  
$\sum_{i=1}^k \lambda_i(S) \leq 0$, where~$S$ is the symmetric 
matrix defined in~\eqref{eq:defmatr}.  Since the eigenvalues of~$S$ are ordered in decreasing
order, we have~$\lambda_k(S) \leq 0$ and thus~$\lambda_j(S) \leq 0$ for any~$j>k$. 
     Hence, $\sum_{i=1}^\ell  \lambda_i(S) \leq 0$, so  condition~\eqref{eq:k_riccati_Pk}
     also holds when we replace~$k$ by~$\ell$. 
     Similarly, we have that~\eqref{cond:k_smallgain_CQ} implies that the same condition also holds when we replace $k$ by any~$\ell > k$, and the same is true for~\eqref{cond:k_smallgain_BQ}. \hfill{\qed}
%%%%%%%%%%%%%%%%%%%%%
\end{pf}
%%%%%%%%%%%%%%%%%%%%%%%%%%%%

%%%%%%%%%%%%%%%%%%%%%%%%%%%
\section{Proof of main result}\label{sec:proof}
%%%%%%%%%%%%%%%%%
This section is devoted to the proof
of Thm.~\ref{thm:lure_suff}. This requires the following 
auxiliary result.

\begin{lem}\label{lem:MN_matrices}
%%%%%
Fix~$M \in \R^{n \times m}$,
$N \in \R^{m \times n}$, and~$k\in\{1,\dots,n\}$.
Then
\[
( - M N - N^T M^T  -  N^T N )^{[k]}\preceq (M M^T) ^{[k]}. 
\]
\end{lem}
\begin{pf}
    %%%
    The  identity
    \[
    M N + N^TM^T = (M^T + N)^T(M^T + N) - MM^T - N^T N 
    \]
    gives
    \[
   Z:=- M M^T  - M N - N^T M^T  -  N^T  N \preceq 0.
   \]
    %%%
    Thus,  $Z$ is symmetric with all (real)
    eigenvalues smaller or equal to zero. Hence, the same properties hold   
      for~$Z^{[k]}$, so
    \[
   Z^{ [ k]}= \left (-  M M^T  - M N - N^T M^T  -  N^T  N \right  )^{[k]}\preceq 0 , 
   \]
   and this completes the proof.~\hfill{\qed}
\end{pf}

We can now prove Theorem~\ref{thm:lure_suff}.
\begin{pf}
%%%%%%%%%%%%%%%%%%%%%%%%%%%5
  Let $R := Q J  Q^{-1} + Q^{-1} J  ^TQ$, with~$J$ defined  in~\eqref{eq:closed_loop_jac}. Then
%%%
    \begin{align*}
    %%%%%%%%%%%%%%%%%%%%
        R^{[k]} &=\left (Q(A - BJ_\phi C)Q^{-1} + Q^{-1}(A - BJ_\phi C)^TQ \right )^{[k]} \\
        &=
        \left (QAQ^{-1} + Q^{-1}A^T Q \right )^{[k]}\\&
        -\left (QBJ_\phi C Q^{-1} + Q^{-1}  C^TJ_\phi^T B^T Q ) \right )^{[k]}.
%%%
\end{align*}
%%%%
    Multiplying~\eqref{eq:k_riccati_Pk} on the left- and on the right-hand side  by~$(Q^{(k)})^{-1}$, and using~\eqref{eq:k_add_coor_trans} gives
%%%%%%%%%%%%%%%
    \begin{multline}\label{eq:kARIcond_Q}
        (QAQ^{-1} + Q^{-1}A^T Q)^{[k]} \preceq \\ -\eta_1 I_r - (QBB^TQ + Q^{-1}C^TCQ^{-1})^{[k]}, 
    \end{multline}
so
   \begin{align}\label{eq:rsofar}
    %%%%%%%%%%%%%%%%%%%%
        R^{[k]} &\preceq  -\eta_1 I_r -\left (QBB^TQ + Q^{-1}C^TCQ^{-1  }\right)^{[k]} \nonumber \\& -\left (QBJ_\phi C Q^{-1} + Q^{-1}  C^TJ_\phi^T B^T Q   \right )  ^{[k]}.
%%%
\end{align}
It follows from Lemma~\ref{lem:MN_matrices} with~$M=Q B J_\phi$ and~$N=CQ^{-1}$ that
\begin{align*}
%%%%%%%%%%%%
 ( - Q B J_\phi  C Q^{-1}  &-  Q^{-1} C^T 
J_\phi^T B^T Q   -  Q^{-1} C^T   CQ^{-1}  )^{[k] }
 \\&\preceq ( Q B J_\phi J_\phi^T B^T Q   ) ^{[k]}. 
  %%%%%
\end{align*}
so
 \begin{align}\label{eq:rkqb}
    %%%%%%%%%%%%%%%%%%%%
        R^{[k]} &\preceq  -\eta_1 I_r 
        + \left (QB (J_\phi J_\phi^T - I_m)B^TQ      
        \right)^{[k]} .
%%%
\end{align}
Also, by Lemma~\ref{lem:MN_matrices} with~$M=Q B  $
and~$N=J_\phi CQ^{-1}$, we have 
%%%%%%%%%%%%%
\begin{align*}
%%%%%%%%%%%%
 ( - Q B J_\phi  C Q^{-1}  &-  Q^{-1} C^T 
J_\phi^T B^T Q   -  Q^{-1} C^T J_\phi^T J_\phi  CQ^{-1}  )^{[k] }
 \\&\preceq ( Q B  B^T Q   ) ^{[k]} ,  
  %%%%%
\end{align*}
and combining this with~\eqref{eq:rsofar} gives
 \begin{align}\label{eq:rk22}
    %%%%%%%%%%%%%%%%%%%%
        R^{[k]} &\preceq  -\eta_1 I_r 
        + \left (Q^{-1}C^T  (J_\phi ^T J_\phi - I_q) C Q^{-1}   
        \right)^{[k]} .
%%%
\end{align}
Thus,
\begin{align*}
   \lambda_{\max}(R^{[k]})  
        &\le -\eta_1  \nonumber\\
        & + \min\{\lambda_{\max}((QB (J_\phi J_\phi^T - I_m)B^TQ)^{[k]}), \nonumber\\
        &\quad\phantom{\min\{}\,\lambda_{\max}(Q^{-1}C^T (J_\phi ^T J_\phi - I_q) C Q^{-1})^{[k]})\} \\
        &\le -\eta_1 -\eta_2, \nonumber
\end{align*}
where the last inequality follows from~\eqref{cond:k_smallgain_CQ} and~\eqref{cond:k_smallgain_BQ}. Since~$ 2\mu_{2,Q^{(k)}}(J^{[k]})  = \lambda_{\max}(R^{[k]}) $, we conclude that if~$\eta_1 + \eta_2 > 0$ then
    the closed-loop system is~$k$-contractive with rate~$ (\eta_1 + \eta_2)/2$ w.r.t. the scaled~$L_2$ norm~$|z|_{2,Q^{(k)}}= |Q^{(k)} z|_2$. This completes the proof of Theorem~\ref{thm:lure_suff}.~\hfill{\qed}

\end{pf}

 %%%%%%%%%%
 \begin{Remark}\label{rem:scalar_P_C}
 %%%%%%%%%%%%%%%%%%%%
Consider the particular case
\[
P=p I_n, \quad  p>0,
\]
i.e.~$Q = p^{1/2} I_n$. Suppose that the~$k$-ARI~\eqref{eq:k_riccati_Pk} 
holds for this~$P$ and for some~$\eta_1>0$.
Suppose that, in addition, 
  \begin{equation}\label{eq:prort}
        \sum_{i=1}^k \sigma_i^2(J_\Phi(t,y))   <k  \text{ for all } t\geq 0,y\in\R^n.
    \end{equation}
We claim that if~$C=I_n$ 
[$B=I_n$] then~\eqref{eq:prort} implies that
\eqref{cond:k_smallgain_CQ} 
[\eqref{cond:k_smallgain_BQ}] holds for some~$\eta_2>0$ and thus the Lurie system is~$k$-contractive. 
To show this, note that if~$C=I_n$ then
\eqref{cond:k_smallgain_CQ} 
becomes 
    \begin{equation*}
        \sum_{i=1}^k \sigma_i^2(J_\Phi(t,y))   \le k-\eta_2 p ,
    \end{equation*}
    and this   always holds for some~$\eta_2>0$ if~\eqref{eq:prort} holds. 
 Similarly, if~$B=I_n$ then~\eqref{cond:k_smallgain_BQ} becomes
  \begin{equation*}
        \sum_{i=1}^k \sigma_i^2(J_\Phi(t,y))   \le k-\eta_2 p^{-1} ,
    \end{equation*}
   and this   always holds for some~$\eta_2>0$ if~\eqref{eq:prort} holds.
%%%
 \end{Remark}
%%%%%%%%%%%%%%%%%%%%%
 
\section{An application: $k$-contraction in a networked~system}\label{sec:networked}
%%%
%%%%%%%%%%%%%%%%%%%%%%
We now apply our main result to analyze  the global behaviour of several models including Hopfield neural networks, a nonlinear opinion dynamics model, and a 2-bus system. The first step is to consider
  a general  networked dynamical system
\begin{equation}\label{eq:net_sys}
    \dot x(t) = -D  x (t)+ W_1f \left (W_2x(t) \right )+v,
\end{equation}
where
  $x \in \Omega \subseteq \R^n$,  $D=\diag(d_1,\dots,d_n)$ is a diagonal matrix, 
  $W_1 \in \R^{n \times m}, W_2 \in \R^{q \times n} $ are matrices of interconnection weights, $v\in\R^n$ is a constant ``offset'' vector, and $f : \R^q \to \R^m$. 
  
  In the context of neural network models, $f$ is typically diagonal, that is,~$q=m$ and
  \[
f(z)=\begin{bmatrix} f_1(z_1)&\dots &f_q (z_q)\end{bmatrix}^T,
  \]
  where the~$f_i$s are the neuron activation functions. More generally, they may represent functions that are bounded or saturated
   and thus non-linear. We assume that the state space~$\Omega$ is convex and that~$f$ is continuously differentiable. Let   
   \begin{equation*}
    J_f(z) = \begin{bmatrix}
        \frac{\partial f_1}{\partial z_1}(z) & \dots & \frac{\partial f_1}{\partial z_q}(z) \\
        \vdots & \ddots & \\
        \frac{\partial f_m}{\partial z_1}(z) & \dots & \frac{\partial f_m}{\partial z_q}(z)
    \end{bmatrix} 
   \end{equation*}
 denote the Jacobian of~$f$. 
 
Intuitively speaking, it is clear that as we take all the~$d_i$s larger the system  becomes ``more stable''.
The next result rigorously 
formalizes this by providing
a sufficient condition for $k$-contraction based on  Theorem~\ref{thm:lure_suff}.
 %%%%%%%%%%%%%%%%%%%%%%
\begin{Theorem}
    \label{thm:net_k_contract}
  Consider~\eqref{eq:net_sys}.  Fix~$k \in [1,n]$, and let
  \be\label{eq:alphak}
  \alpha_k:=\frac{1}{k}\min\left\{ d_{i_1}+\dots+d_{i_k}\st 1\leq i_1<\dots<i_k\leq n \right \}.
  \ee
If $ \alpha_k>0 $ and 
    \begin{equation}\label{cond:net_k_smallgain}
     \|J_f(W_2x)\|_2^2 \sum_{i=1}^k \sigma_i^2(W_1)\sigma_i^2(W_2) < \alpha_k^2 k  \text{ for all } x \in \Omega,
    \end{equation}
    then~\eqref{eq:net_sys} is $k$-contractive. 
    Furthermore, if these conditions hold for~$k=2$ then every bounded trajectory of~\eqref{eq:net_sys} converges to an equilibrium point (which is not necessarily unique).
%%%%%%%%%
\end{Theorem}
%%%%%%%%%%
\begin{Remark}
Note that condition~\eqref{cond:net_k_smallgain} does not require to  explicitly compute any~$k$-compounds. This is useful, as for  a matrix~$A\in\R^{n\times n}$ the~$k$-compounds have dimensions~$\binom{n}{k}\times\binom{n}{k}$, and this may be quite large (see also~\cite{Dalin2022Duality_kcont}).
The condition~$\alpha_k>0$ is equivalent to requiring that 
the sum  of every~$k$ eigenvalues of~$D$ is positive. For~$k=1$, this amounts to requiring that~$D$ is a positive diagonal matrix, but for~$k>1$  some of the~$d_i$s may be negative, as long as the sum of every~$k$ of the~$d_i$s is positive.
%%
%%%%%%%%%%%%%%
\end{Remark}
%%%%%%%%
\begin{pf}
%%%%%%%%%%%%%%%%%%%%%%%%%%
The proof is based on Theorem~\ref{thm:lure_suff}. We first represent~\eqref{eq:net_sys} as a Lurie system. By~\eqref{cond:net_k_smallgain}, there exists~$\gamma\in\R$ satisfying 
\be\label{eq:gammprop}
    0<\gamma < \alpha_k \text{ and } \|J_f(z)\|_2^2 \sum_{i=1}^k \sigma_i^2(W_1) \sigma_i^2(W_2)< \gamma^2 k.
\ee
We can represent~\eqref{eq:net_sys} as 
the interconnection of the LTI system with $(A,B,C)=(-D, \gamma I_n, I_n)$ and the nonlinearity~$\Phi(y) := -\gamma^{-1}W_1f(W_2 y) -\gamma^{-1}v$, that is,
\begin{align}\label{eq:lui}
    \dot x&= - D  x + \gamma u,\nonumber\\
    y&=x,\nonumber\\
    u&= \gamma^{-1} W_1f(W_2  y  )+\gamma^{-1}v.
    %%%%
\end{align}
For this Lurie system, there exist~$Q \succ 0$ with~$P=QQ$ and~$\eta_1>0$ such that the~$k$-ARI~\eqref{eq:k_riccati_Pk} holds if and only if
\begin{equation}\label{eq:k_riccati_net}
 - P^{(k)}D^{[k]}
     -D^{[k]} P^{(k)}+ Q^{(k)}(\gamma^2 P + P^{-1})^{[k]}Q^{(k)} \prec  0.
\end{equation}

Taking~$P = p I_n$, with~$p>0$,   gives
\begin{equation}\label{eq:k_riccati_scalar}
  \left(  -2 D^{[k]}+
    (  \gamma^2 p + p^{-1}) k  I_r \right  ) p^k \prec   0.
\end{equation}
By  definition,~$ \alpha_k k$ 
is a lower bound of   the diagonal  entries of~$D^{[k]}$. Thus, Eq.~\eqref{eq:k_riccati_scalar}
will hold for any~$p>0$ such that
\[
   -2\alpha_k+ 
  \gamma^2 p   + p^{-1}        < 0, 
\]
and this indeed  admits a solution~$p>0$ since~$\alpha_k > 0$ and~$\gamma < \alpha_k$. We conclude that there exists a matrix~$P = p I_n$, with~$p>0$, and a scalar~$\eta_1 > 0$ for which the~$k$-ARI~\eqref{eq:k_riccati_Pk} holds.

We now show that~\eqref{cond:net_k_smallgain} implies that~\eqref{cond:k_smallgain_CQ} holds for some~$\eta_2>0$.
Since~$P=pI_n$ and~$C=I_n$, we may apply the result in Remark~\ref{rem:scalar_P_C}. Recall that for any~$A\in\R^{m\times p},B\in\R^{p\times n}$, we have
\be\label{eq:sigmai2}
    \sum_{i=1}^k \sigma_i^s(AB) \leq
    \sum_{i=1}^k (\sigma_i(A)\sigma_i(B))^s  
\ee
for any $k\in[1,\min\{m,p,n\}]$, $s>0$
\citep[Thm.~3.3.14]{Horn1991TopicsMatrixAna}. Consider
\begin{align*}
    \sum_{i=1}^k \sigma_i^2(J_\Phi)
& = \sum_{i=1}^k \sigma_i^2 (-\gamma^{-1} W_1 J_f W_2)\\
&\leq\gamma^{-2}  \sum_{i=1}^k \sigma_i^2 (  W_1 J_f )  \sigma_i^2 (  W_2  ) \\
&    \leq \gamma^{-2} \sigma_1^2 (    J_f )  \sum_{i=1}^k \sigma_i^2 (  W_1   )  \sigma_i^2 (  W_2  ) \\
&<k,
\end{align*}
where the first two inequalities follows from~\eqref{eq:sigmai2}, 
and the third from~\eqref{eq:gammprop}. 
We conclude that the sufficient condition~\eqref{eq:prort} holds,
and  Theorem~\ref{thm:lure_suff} implies that~\eqref{eq:net_sys} is $k$-contractive.

Suppose now that~\eqref{cond:net_k_smallgain} holds with~$k=2$. Then~\eqref{eq:net_sys} is 2-contractive. If in addition~$f$ is uniformly bounded, then all the  trajectories of~\eqref{eq:net_sys} are bounded, and by known  results on time-invariant  2-contractive systems~\citep{li1995} we then have that every trajectory    converges to an equilibrium point. This completes the proof of Theorem~\ref{thm:net_k_contract}.~\hfill{\qed}
%%%%%%%%
\end{pf}

%%%%%%%%%%%%%%%%%%%%%%

 \begin{Remark}
     In the special case where~$D=\alpha I_n$,   the  networked dynamical system becomes
\begin{equation}\label{eq:net_sys_DI}
    \dot x = -\alpha x + W_1f(W_2x),
\end{equation}
and the sufficient condition for~$k$-contraction is 
  %%%%%%%%%%%%%%%%%%%%%%
    \begin{equation}\label{cond:net_k_smallgain_special}
    %%%%%%%%
  \alpha>0 \text{ and }  \|J_f(W_2x)\|_2^2 \sum_{i=1}^k \sigma_i^2(W_1)\sigma_i^2(W_2) < \alpha^2 k  ,
    \end{equation}
%%%%%%%%%%
  for all $x \in \Omega$. 
   Note also that if either~$f=0$ or~$W_1=0$ or~$W_2=0$ then~\eqref{cond:net_k_smallgain_special} holds for~$k=1$ (and thus for any~$k\in [1,n]$). This is reasonable,  as in this case we have~$\dot x=-\alpha x$, and this is indeed $k$-contractive for any~$k\geq 1$. 
%%%%%%%%%%%%%%
 \end{Remark}

We now apply Theorem~\ref{thm:net_k_contract}
to three  specific models:  a Hopfield neural network,
a nonlinear opinion dynamics system, and a 2-bus power system. All these  application are typically multi-stable, that is, they include more than a single equilibrium point, and thus are not contractive (i.e., not $1$-contractive) w.r.t. any norm. However,  our results may still be applied to prove~$k$-contraction, with~$k>1$. 

\subsection{2-Contraction in Hopfield neural networks}
%%%%%%%%%%%%%%%%%%%%%%%%%%%%%%%%%%%%%%%%%%%%%%%
A particular example of a networked system in the form~\eqref{eq:net_sys} is the well-known Hopfield neural network~\citep{hopfield_net}:
\begin{equation}\label{eq:hopfield}
    \dot x = -\alpha x + W f(x).
\end{equation}
The stability of this model has been studied extensively.   \cite{cohen_NN} used a Lyapunov function  to 
prove then when~$W$
is symmetric and the system is competitive  each trajectory converges to the set of equilibria.
\cite{cont_hopfield} analyzed the stability of~\eqref{eq:hopfield} 
 using contraction theory. However, 
the  system is often multistable, and  thus not contractive (i.e., not 1-contractive) w.r.t. any norm.
For example,~\citep{multistableNNs} found conditions guaranteeing that an
$n$-dimensional Hopfield network with logistic activation functions has~$3^n$ equilibrium points. 
Moreover, Hopfield networks are often used as
  associative memories, where each equilibrium corresponds to a stored pattern (see, e.g.,~\cite{krotov2016}),
  so   multistability is in fact  a desired property.
  
  Here we consider the typical choice of using $\tanh(\cdot)$ as the activation function, i.e.,  taking\be\label{Eq:tabhh}
  f(x) = \begin{bmatrix}
  \tanh(x_1)&\dots&\tanh(x_n)
  \end{bmatrix}^T.
\ee
Note that this implies that~$\| J_f(x)\|_2^2 \leq 1$ for any~$x\in\R^n$.

\begin{Corollary}
\label{coro:hopfield}
Consider the Hopfield network defined 
by~\eqref{eq:hopfield} and~\eqref{Eq:tabhh}.
    If 
\begin{align}\label{eq:hop1c}
 \sigma_1 (W)   < \alpha  
    \end{align}
then the   network is contractive. If
    \begin{align}\label{eq:rhp}
  \sqrt{ \sigma_1^2 (W)   + \sigma_2^2(W)} < \sqrt{2}\alpha  
    \end{align}
    then the network is~$2$-contractive and every solution 
    converges to an equilibrium point.
\end{Corollary}
\begin{pf}
First, note that it follows from~\eqref{eq:hopfield} and~\eqref{Eq:tabhh} that every solution of the Hopfield network is bounded. 
Second,  note that~\eqref{eq:hopfield} is a special case of~\eqref{eq:net_sys_DI} with~$W_1=W$ and~$W_2=I_n$, so we can
apply  Theorem~\ref{thm:net_k_contract} to the Hopfield network model. In this case,~\eqref{eq:alphak}
gives~$\alpha_k=\alpha$ for all~$k$,    so~\eqref{cond:net_k_smallgain}
becomes~$\alpha>0$ and~$
   \sum_{i=1}^k \sigma_i^2(W )  < \alpha^2 k$. 
  In the particular case~$k=2$ this is equivalent to~\eqref{eq:rhp}, and     this implies  that every bounded solution    converges to an equilibrium point.~\hfill{\qed}
%%%%%%%%%%%%%%%%
\end{pf}

\begin{figure}
    \centering
    \includegraphics[width=\linewidth]{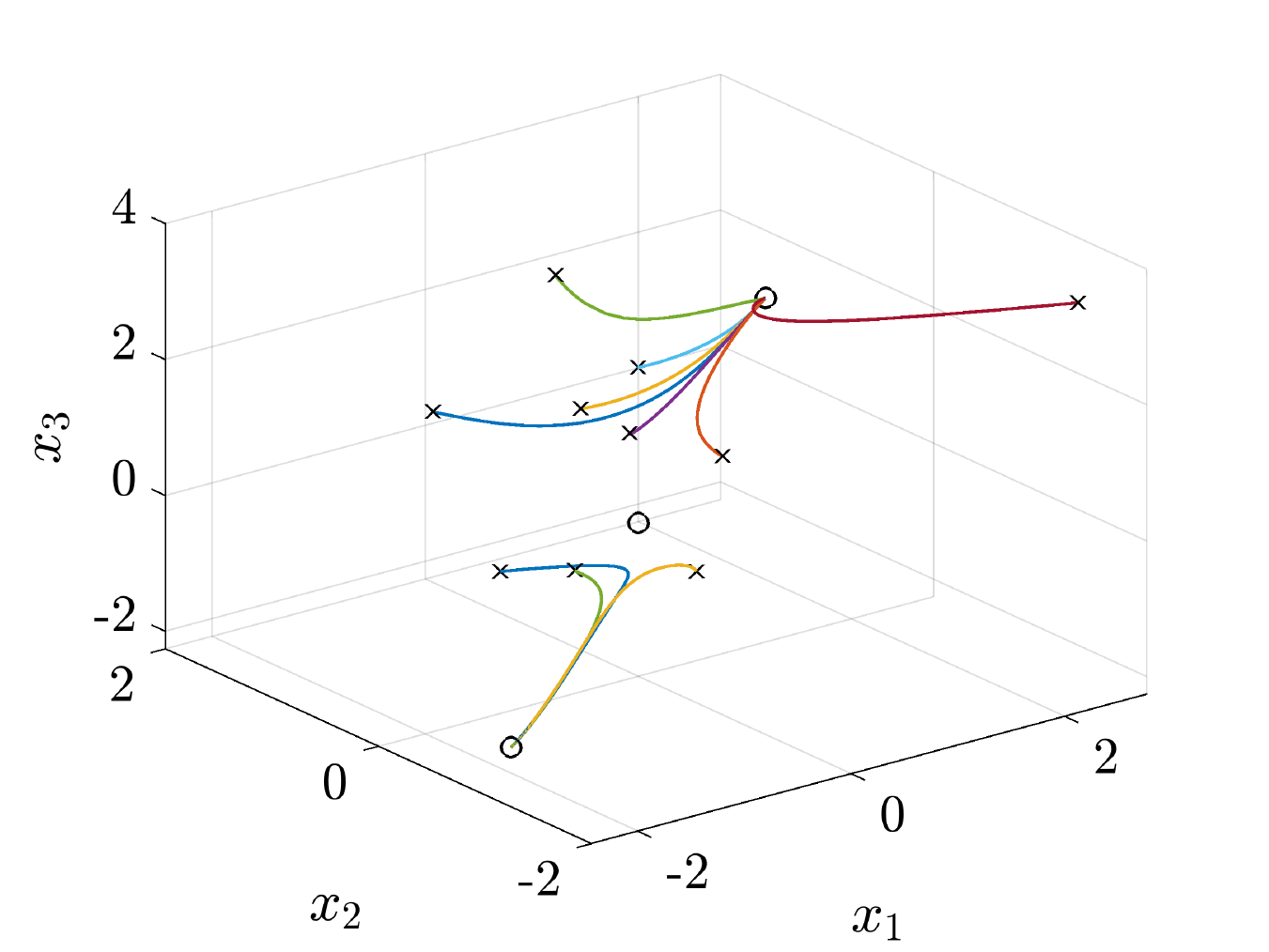}
    \caption{Several trajectories of the Hopfield network described in Example~\ref{exa:hopfield}. The equilibrium points of the system are marked by circles. Initial conditions are marked with crosses.  }
    \label{fig:hopfield}
\end{figure}

The next example demonstrates that  Corollary~\ref{coro:hopfield} may be used to analyze the case where 
the network is multi-stable, and thus  it is certainly not contractive (i.e., not~$1$-contractive) w.r.t. any norm. We consider the case~$n=3$, as then we can plot the system trajectories. 
%%%%%%%%%%%%%%%%%%%%
\begin{Example}\label{exa:hopfield}
%%%%%%%%%%%%%%%%%%%%
    Consider a Hopfield network with~$3$ neurons and
    \begin{equation*}
        W = \begin{bmatrix}
            0 & 1 & 1 \\
            0 & 0 & 1 \\
            1 & 0 & 0
        \end{bmatrix}.
    \end{equation*}
    Note that~$W$ is not symmetric.
    In this case, $\sigma_1^2(W) = (3 + \sqrt{5})/2 \approx 2.618$ and $\sigma_2^2(W) = 1$. Corollary~\ref{coro:hopfield} implies that the network is contractive when
    \[
    \alpha>(3 + \sqrt{5})/2\approx 2.618,
    \]
    and 2-contractive when 
    \[
    \alpha > \sqrt{\frac{5 + \sqrt{5}}{4}}\approx 1.345.
    \]
    Consider the case~$\alpha = 1.5$. Then the network has at least three  equilibrium points, namely, 
    $e^1=0$, $e^2 \approx \begin{bmatrix}2.435&1.243&1.3870\end{bmatrix}^T$ and $e^3 = -e^2$. Thus the network is multistable and so it is not $1$-contractive with respect to any norm. Furthermore, since condition~\eqref{eq:rhp} holds, the system is 2-contractive. Fig.~\ref{fig:hopfield} shows several trajectories of the system with the described parameters. It may be seen that as expected, every solution converges to an equilibrium point. 
%%%%%%%%%%%%
\end{Example}

\begin{comment}
MATLAB CODE FOR HOPFIELD NETWORK FIGURE:

set(0,'defaulttextinterpreter','latex');
set(0,'defaultAxesTickLabelInterpreter','latex');
set(0,'defaultfigurecolor',[1 1 1]);
set(0,'defaultaxesfontsize',16);
set(0,{'DefaultAxesXColor','DefaultAxesYColor','DefaultAxesZColor','DefaultTextColor'},...
     {'k','k','k','k'});

n = 3;           % number of neurons
W = [0,1,1;0,0,1;1,0,0];%1/n*ones(n); % interconnection network
alpha = 0.71;        % self-feedback gain

% Try to find three equilibrium points 
e1 = fsolve(@(x) -alpha*x + W*tanh(x), [0;0;0]);
e2 = fsolve(@(x) -alpha*x + W*tanh(x), [1;1;1]);
e3 = -e2;

% Plot several trajectories starting from random initial conditions
figure();
for i = 1:10
    [~,x] = ode45(@(t,x) -alpha*x + W*tanh(x), [0 15], randn(3,1));
    plot3(x(:,1), x(:,2), x(:,3), 'LineWidth', 1); hold on;
    scatter3(x(1,1),x(1,2),x(1,3),'xk', 'LineWidth', 1);
end
% Add the three equilibrium points the figure
E = [e1,e2,e3];
scatter3(E(1,:),E(2,:),E(3,:),'ok', 'LineWidth', 1);
grid on;
xlabel('$x_1$', 'Interpreter', 'latex');
ylabel('$x_2$', 'Interpreter', 'latex');
zlabel('$x_3$', 'Interpreter', 'latex');

\end{comment}

%%%%%%%%%%%%%%%%%%%%%%%%%%%%%%%%%%%%%%%%%
\subsection{An application to a nonlinear opinion dynamics model}
%%%%%%%%%%%%%%%%%%%%%%%%%%%%%%%%%%%
In this section, we consider the nonlinear opinion dynamics model recently
proposed and analyzed by \cite{opinion_leonard}. For the two-option case, the model is given by
\be\label{eq:opin}
\dot x_i(t) = -d_i x_i + u_i f \left(    \sum_{j =1 }^n a_{ij} x_j (t) \right)+b_i,\quad i\in[1,n] ,
\ee
where~$d_i >0$, and~$f:\R\to\R$ is an odd saturating function. 
Here~$x_i$ represents the opinion of agent~$i$, the term~$ \sum_{j =1 }^n a_{ij} x_j$ is the cue obtained from all the  agents that communicate over a network with weights~$a_{ij}$, the term~$-d_i x_i$ represents a ``forgetting term'', the parameter~$u_i$ determines how ``attentive'' is agent~$i$ to the opinions of the agents, and~$b_i \ge 0$ is a constant  offset (``bias'') term.   
 
\cite{opinion_leonard} showed that the nonlinear function~$f$ in the model introduces many behaviours that cannot be captured using linear consensus systems. In particular, for the homogeneous case where $d_i \equiv d, u_i \equiv u \ge 0, a_{ii} \equiv a, a_{ij} \ge 0$, and $A$ irreducible, the model goes through a pitchfork bifurcation as~$u$ grows larger: that is, if~$u$ is larger than a certain threshold depending on the topology of the interconnection network, then the model has multiple equilibrium points, several of which are stable. However,  \cite{opinion_leonard} only studied  local stability. In this section, we use Theorem~\ref{thm:net_k_contract} to study $k$-contraction in this model, which for the case of $k=2$ will prove global asymptotic stability.

To apply our results, note that~\eqref{eq:opin} can be written as in~\eqref{eq:net_sys} 
with~$D=\diag(d_1,\dots,d_n)$,  $W_1=\diag(u_1,\dots,u_n)$,
$W_2=A=\{a_{ij}\}_{i,j=1}^n$, and~$v=b=\begin{bmatrix}b_1&\dots&b_n
\end{bmatrix}^T$. 
Applying Theorem~\ref{thm:net_k_contract} yields the following result.
\begin{Corollary}\label{coro:opinion_k_cont}
%%%%%%%%%%%
    Consider~\eqref{eq:opin} and assume  without loss of generality that the state-variables are ordered such that~$u_1^2\geq\dots\geq u_n^2$.  Fix~$k \in [1,n]$, and let
    \[
      \alpha:=\frac{1}{k}\min\left\{ d_{i_1}+\dots+d_{i_k}\st 1\leq i_1<\dots<i_k\leq n \right \}  .
    \]
    If $\alpha>0 $ and 
    \begin{equation}\label{cond:net_k_smallgain_opinion}
        \|J_f(Ax)\|_2^2 \sum_{i=1}^k u_i^2 \sigma_i^2(A) < \alpha^2 k \text{ for all } x \in \Omega 
    \end{equation}
    then~\eqref{eq:opin} is $k$-contractive. 
    Furthermore, if~$f$ is uniformly bounded and~\eqref{cond:net_k_smallgain_opinion} holds with $k=2$  then every trajectory of~\eqref{eq:opin} converges to an equilibrium point (which is not necessarily unique).
%%%%%%%%%%% 
\end{Corollary}

\begin{figure}
    \centering
    \includegraphics[width=\linewidth]{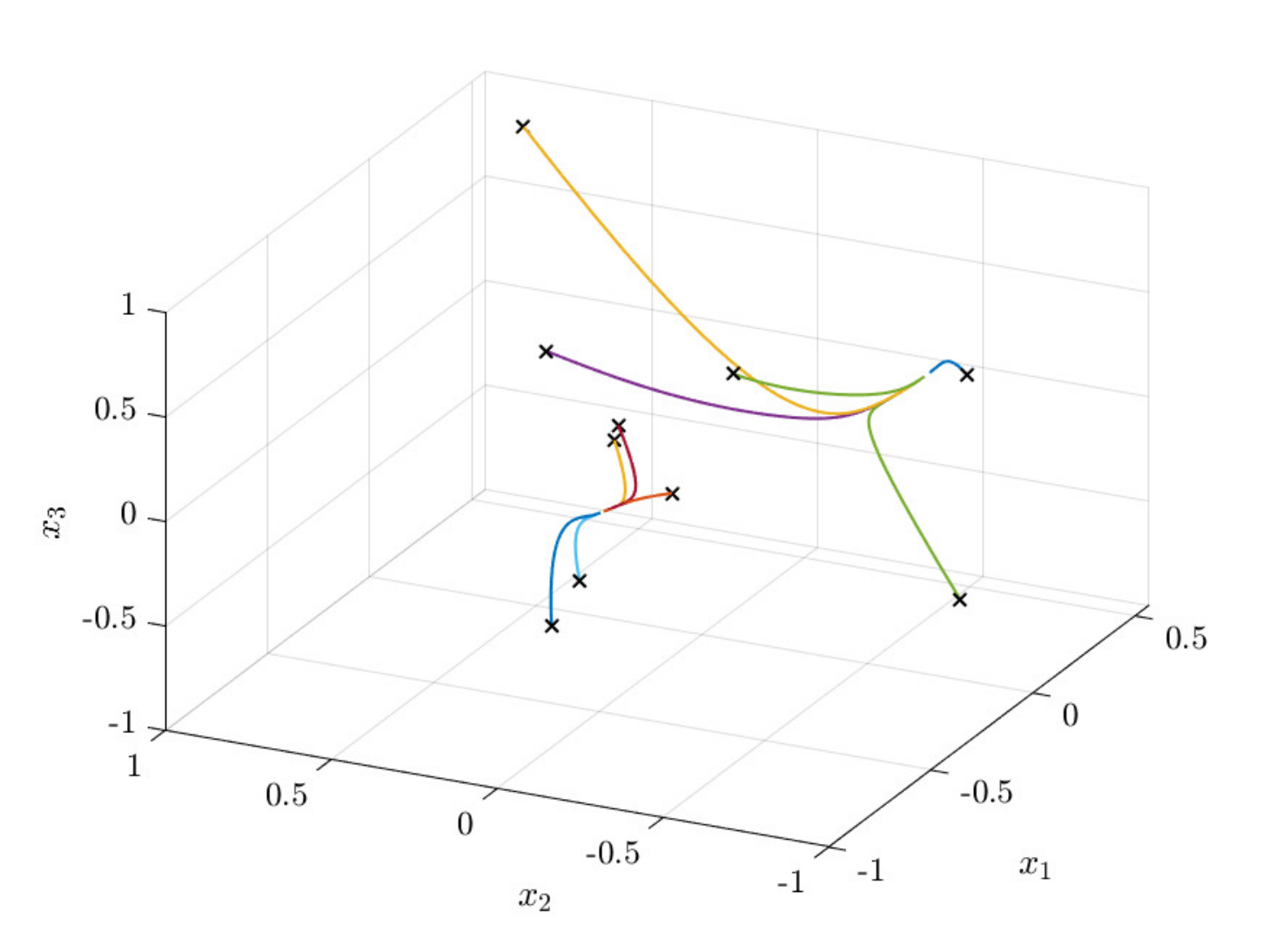}
    \caption{Numerical simulation of several trajectories of the opinion dynamics model in Example~\ref{exa:opinion} with $u=0.5$. Initial conditions are marked with crosses.}
    \label{fig:opinion}
\end{figure}

\begin{Example}\label{exa:opinion}
%%%%%%%%
Consider~\eqref{eq:opin} with $n=3$ agents,
$D = I_3$, $W_1 = u I_3$, with~$u > 0$, $b = \begin{bmatrix} 0.2 & 0 & -0.2 \end{bmatrix}^T$,
  connection matrix
    \begin{equation*}
      A = \begin{bmatrix}
            1 & 0 & 0 \\
            0 & 1 & 0 \\
            0 & 0 & 1
        \end{bmatrix} - \begin{bmatrix}
            0 & 1 & 0 \\
            1 & 0 & 1 \\
            0 & 1 & 0
        \end{bmatrix},
    \end{equation*}
    and $f$ as in~\eqref{Eq:tabhh}. It then follows from Corollary~\ref{coro:opinion_k_cont} that the system is $k$-contractive if
    \begin{equation}
        u^2\sum_{i=1}^k \sigma_i^2(A) < k.
    \end{equation}
    In this case,
    $\sigma_1^2(A)=3+2\sqrt{2} $, 
$\sigma_2^2(A)= 1$, 
and~$\sigma_3^2(A)= 3 - 2\sqrt{2} $,  so
 the system is $1$-contractive for $u < (1 + \sqrt{2})^{-1} \approx 0.414$, it is    $2$-contractive for $u < \sqrt{ \frac{2}{4+2\sqrt{2}}} \approx 0.541$, and $3$-contractive for $u < \sqrt{ \frac{3}{7}} \approx 0.655$. Several trajectories of this model with~$u=0.5$ (for which the system is 2-contractive) are shown in Fig.~\ref{fig:opinion}. It may be seen that there exist at least two equilibrium points, so  the system is indeed not 1-contractive for these parameter values, and every trajectory converges to an equilibrium. Using~\cite[Corollary IV.1.2]{opinion_leonard}, it can be verified that the bifurcation for this example occurs at $u^* = (1+\sqrt{2})^{-1}$, which is exactly the point at which the system transitions from 1-contraction to 2-contraction according to Thm.~\ref{thm:net_k_contract}. Hence, in this case, Thm.~\ref{thm:net_k_contract} is exact rather than conservative.
\end{Example}

\begin{comment}
MATLAB CODE FOR BOUND CALCULATIONS AND NUMERICAL SIMULATION:

% Set defaults for nicer plots
set(0, 'defaulttextinterpreter', 'latex');
set(0, 'defaultAxesTickLabelInterpreter','latex');
set(0, 'defaultfigurecolor', [1 1 1]);
set(0, 'defaultaxesfontsize', 11);
set(0, {'DefaultAxesXColor', 'DefaultAxesYColor', 'DefaultAxesZColor', 'DefaultTextColor'}, ...
       {'k', 'k', 'k', 'k'});

% System parameters, taken from Fig. 4 in "Non-linear opinion dynamics with
% tunable sensitivity" by Bizyaeva et al.
n = 3;
D = eye(n);
gamma = -1;
alpha = 1;
W2 = [alpha, gamma, 0     ;
      gamma, alpha, gamma ;
      0    , gamma, alpha];
v = [0.2;0;-0.2];

% Calculate bounds on u for k-contraction, with k = 1,2,3
E = svd(W2);
for k = (1:3)
    sqrt(k / sum(E(1:k).^2))
end

% Plot several trajectories starting from random initial conditions
figure();
for i = 1:10
    [~,x] = ode45(@(t,x) -x + W1*tanh(W2*x) + v, [0 15], 2*rand(3,1) - 1);
    plot3(x(:,1), x(:,2), x(:,3), 'LineWidth', 1); hold on;
    scatter3(x(1,1),x(1,2),x(1,3),'xk', 'LineWidth', 1);
end
grid on;
xlabel('$x_1$', 'Interpreter', 'latex');
ylabel('$x_2$', 'Interpreter', 'latex');
zlabel('$x_3$', 'Interpreter', 'latex');
\end{comment}

\subsection{An application to power systems}
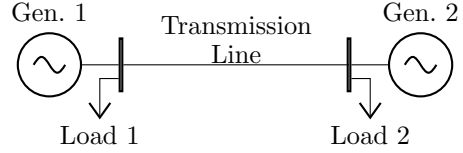
\begin{figure}[t]
    \centering
    \begin{circuitikz}[european,busbar/.style = {draw, rectangle, thick, minimum height=2em, minimum width=0.1em, inner sep=0em}]

    \def\normalcoord(#1){coordinate(#1)}
    \def\showcoord(#1){node[circle, red, draw, inner sep=1pt, pin={[red, overlay, inner sep=0.5pt, font=\tiny, pin distance=0.1cm, pin edge={red, overlay}]45:#1}](#1){}}
    \let\coord=\normalcoord
    
    \node[busbar](bus1){};
    \draw (bus1.west) to[short] ++(-0.5,0) node[label={left:Gen. 1},vsourcesinshape,anchor=north,rotate=-90](SG1){};
    \draw ($(bus1.west)!0.5!(bus1.south west)$) to[short] ++(-0.25,0) to[short] ++(0,-0.1) node[vee,label={[label distance=0.4 cm]below:Load 1}](){};
    
    \draw (bus1.east) to[short,l={\parbox{2cm}{\centering Transmission\\Line}},label distance=0.1cm] ++(3,0) node[busbar](bus2){};
    
    \draw (bus2.east) to[short] ++(0.5,0) node[label={right:Gen. 2},vsourcesinshape,anchor=north,rotate=90](SG2){};
    \draw ($(bus2.east)!0.5!(bus2.south east)$) to[short] ++(0.25,0) to[short] ++(0,-0.1) node[vee,label={[label distance=0.4 cm]below:Load 2}](){};
    
    \end{circuitikz}
    \caption{Schematic description of the 2-bus power system. A synchronous generator (depicted  as an AC source) and a constant power load (indicated by an  arrow) are connected to a each bus locally, and the two buses are connected to each other over a transmission line.}
    \label{fig:2bus}
\end{figure}

%%%%%%%%%%%%%%%%%%%%%%%%%%%%%%%
We now use our results to provide a global stability result for a power system consisting of two interconnected synchronous generators (see Fig.~\ref{fig:2bus}) based on the so-called Network-Reduced Power System (NRPS) model~\citep{Sauer1998power}. A useful approach for analysing the stability of the NRPS model, that is based on singular perturbation theory,
was first proposed by~\cite{Dorfler2012kuramoto}, and recently extended by~\cite{Weiss2019StabilityNRPS}. In this approach, the NRPS is related to a Nonuniform Kuramoto  model, where the stability can be studied analytically. However, since the approach is based on singular perturbations, it typically yields a highly conservative bound on the inertia of the system. In this section, we focus on the case of a system with two generators and derive a sufficient condition for 2-contractivity, which implies that all bounded trajectories converge to an equilibrium point.

%(denoted by \begin{circuitikz}[baseline=0.1ex,scale=0.75, transform shape]\node [vee] {};\end{circuitikz})
%\draw (0,0) -- (.1,0) node [ac source,anchor=west,name=s] {} (s.east) -- +(.1,0);

Following the network reduced power system model, the system under study is described by
\begin{align}\label{eq:2bus}
    M_1 \dot \omega_1(t) &= p_1 -R_1 \omega_1(t) - a \sin(\delta(t) + \varphi), \nonumber\\
    M_2 \dot \omega_2(t) &= p_2 -R_2 \omega_2(t) + a \sin(\delta(t) - \varphi), \nonumber\\
    \dot \delta(t) &= \omega_2(t) - \omega_1(t),
\end{align}
where~$\omega_1,\omega_2 :\R_+\to  \R$ are the rotor rotational frequencies of the two generators,~$\delta : \R_+\to  \R$ is the phase angle of the second generator in reference to the first,~$R_i > 0, i=1,2$, are the damping coefficients,~$M_i >0, i=1,2$, are the inertia constants, $p_1,p_2 > 0$ are the constant power consumption at each bus, and~$a > 0$ and~$\varphi \in (-\pi/2, \pi/2)$ describe the nominal voltages of the generators and the admittance of the transmission line (see~\cite{Weiss2019StabilityNRPS} for a detailed derivation of this model).

\begin{Corollary}\label{coro:2bus_2cont}
    Suppose that~$a > \max\{M_1,M_2\}$. If
    \begin{equation}\label{cond:2bus_2cont}
        3a^2\left(1 + |\cos(2\varphi)|\right) < \frac{\displaystyle \min_i\{M_i\}}{\displaystyle \max_i \{M_i\}}\min_{i  } \frac{R_i^2}{2},
    \end{equation}
    then~\eqref{eq:2bus} is 2-contractive.
\end{Corollary}

\begin{pf}
    Our proof is based on Theorem~\ref{thm:net_k_contract}. First note that
    we can write~\eqref{eq:2bus} as the networked system~\eqref{eq:net_sys} with: 
     $x=\begin{bmatrix} 
    \omega _1&\omega _2&\delta  \end{bmatrix}^T $,
$D = \diag(R_1/M_1,R_2/M_2,0)$, $v=\begin{bmatrix}\frac{p_1}{M_1} &\frac{p_2}{M_2}&0\end{bmatrix}^T$, $W_1 = \diag(-a/M_1,a/M_2,1)$,
    $ 
        W_2 = \begin{bmatrix}
            0 & 0 & 1 \\
            -1 & 1 & 0
        \end{bmatrix},
$ so that~$W_2x=\begin{bmatrix} 
 \delta &  \omega_2-\omega_1  \end{bmatrix}^T$, 
  and     \[  f(z) = \begin{bmatrix}
            \sin(z_1 + \varphi) \\
            \sin(z_1 - \varphi) \\
            z_2
        \end{bmatrix}.
    \]
    Thus,~\eqref{eq:alphak} gives
    \[
    \alpha_2=\frac{1}{2} \min_i \left \{\frac{R_i}{M_i}  \right\} ,
    \] 
    and 
    \begin{equation*}
        J_f(z) = \begin{bmatrix}
            \cos(z_1 + \varphi) & 0 \\
            \cos(z_1 - \varphi) & 0 \\
            0 & 1
        \end{bmatrix},
    \end{equation*}
      so 
      \begin{align*}
          (J_f(z))^TJ_f(z) & = \begin{bmatrix}
              \cos^2(z_1+\varphi)+\cos^2(z_1-\varphi) & 0\\
              0 &1
          \end{bmatrix}\\
          &= 
          \begin{bmatrix}
             1+\cos(2z_1)\cos(2\varphi) & 0\\
              0 &1
          \end{bmatrix},
      \end{align*}
      and thus
      \begin{align*}
      \|J_f(z)\|_2^2&=\lambda_{\max}\left((J_f(z))^TJ_f(z)\right)\\& \le 1 + |\cos(2\varphi)| .
      %%%%%%%%%%%
      \end{align*}
      Furthermore, the ordered singular values of~$W_1$ are
    \[
       \frac{a}{\min\{M_1,M_2\}},\; \frac{a}{\max\{M_1,M_2\}}, \;1  ,
    \]
    and the singular values of~$W_2$ are $ \sqrt{2},\;1 $. Substituting all these
    values in~\eqref{cond:net_k_smallgain} gives 
    \begin{align*}
        \| & J_f (W_2x)\|_2^2 \sum_{i=1}^2 \sigma_i^2(W_1)\sigma_i^2(W_2) \\
             &\leq \left(1 + |\cos(2\varphi)|\right) \left( \frac{2a^2}{(\min\{M_i\})^2} + \frac{a^2}{(\max\{M_i\})^2} \right) \\
             &\le \frac{3a^2}{(\min\{M_i\})^2}\left(1 + |\cos(2\varphi)|\right) \\
             &< \frac{1}{(\max\{M_i\})^2}\min_{i } \frac{R_i^2}{2} \\
             &\le \min_{i  } \frac{R_i^2}{2M_i^2}\\
             &= 2\alpha_2^2,
    \end{align*}
    where we used~\eqref{cond:2bus_2cont} in the last inequality. Therefore,~\eqref{cond:net_k_smallgain} holds with $k=2$.~\hfill{\qed}
\end{pf}
%%%%%%%%%
%%%%%%%%
To relate condition~\eqref{cond:2bus_2cont} to the results of~\cite{Weiss2019StabilityNRPS}, note that the system will always be 2-contractive if the damping coefficients are large enough or if the inertia constants are small enough.

%%%%%%%%%%%%%%%%%%%
%%%%%%%%%%%%%%%%%%%%%
%%%%%%%%%%%%%%%%%%%%%%%%

%%%%%%%%%%%%%%%%%%%%%%%%%%%%%
\section{Conclusion}
%%%%%%%%%%%%%%%%%%%%%%%%%%%%%%%%%
We derived a sufficient condition for~$k$-contraction of Lurie systems. For~$k=1$, this reduces to the standard small gain sufficient condition for contraction. However, often Lurie systems admit more than a single  equilibrium point, and are thus not contractive (that  is, not~$1$-contractive) with respect to
any norm. 

Our condition may still be used to guarantee a well-ordered behaviour of the closed-loop system. For example,     establishing that  a  time-invariant system is~$2$-contractive implies that 
any bounded solution converges  to an equilibrium, that is not necessarily unique. Such a property is important, for example, in dynamical models of associative memories, where every equilibrium corresponds to a stored memory. 

Our results suggest several possible research directions. 
First, an important advantage of ARIs is that they are equivalent to linear  matrix inequalities and there exist efficient numerical algorithms for  solving 
them. An interesting   question is whether this remains  true for the k-ARIs developed here. 
Second,  several criteria 
for the asymptotic stability of a Lurie system, e.g.   the Popov criterion and the circle criterion  can be stated 
using the transfer function of the linear subsystem. It may be of interest to relate the conditions in Theorem~\ref{thm:lure_suff}
to the  transfer function of a linear system with $k$-compound matrices.

{\sl Acknowledgments.} We are grateful to the anonymous reviewers for many helpful comments.
We thank Rami Katz and Francesco Bullo
for useful comments on a previous version of this paper.

%%%%%%%%%%%%
\bibliographystyle{abbrvnat}

\begin{thebibliography}{43}
\providecommand{\natexlab}[1]{#1}
\providecommand{\url}[1]{\texttt{#1}}
\expandafter\ifx\csname urlstyle\endcsname\relax
  \providecommand{\doi}[1]{doi: #1}\else
  \providecommand{\doi}{doi: \begingroup \urlstyle{rm}\Url}\fi

\bibitem[Aminzare and Sontag(2014)]{sontag_cotraction_tutorial}
Z.~Aminzare and E.~D. Sontag.
\newblock Contraction methods for nonlinear systems: A brief introduction and
  some open problems.
\newblock In \emph{{Proc.\ 53rd IEEE Conf. on Decision and Control}}, pages
  3835--3847, Los Angeles, CA, 2014.

\bibitem[Andrieu and Tarbouriech(2019)]{Andrieu2019LMIContract}
V.~Andrieu and S.~Tarbouriech.
\newblock {LMI} conditions for contraction and synchronization.
\newblock \emph{IFAC-PapersOnLine}, 52\penalty0 (16):\penalty0 616--621, 2019.
\newblock 11th IFAC Symposium on Nonlinear Control Systems (NOLCOS 2019).

\bibitem[Bar-Shalom et~al.(2023)Bar-Shalom, Dalin, and
  Margaliot]{comp_long_tutorial}
E.~Bar-Shalom, O.~Dalin, and M.~Margaliot.
\newblock Compound matrices in systems and control theory: a tutorial.
\newblock \emph{Math. Control Signals Systems}, 2023.

\bibitem[Bizyaeva et~al.(2023)Bizyaeva, Franci, and Leonard]{opinion_leonard}
A.~Bizyaeva, A.~Franci, and N.~E. Leonard.
\newblock Nonlinear opinion dynamics with tunable sensitivity.
\newblock \emph{IEEE Trans.\ Automat.\ Control}, 68\penalty0 (3):\penalty0
  1415--1430, 2023.

\bibitem[Bullo(2022)]{bullo_contraction}
F.~Bullo.
\newblock \emph{Contraction Theory for Dynamical Systems}.
\newblock Kindle Direct Publishing, 2022.
\newblock URL \url{http://motion.me.ucsb.edu/book-ctds}.

\bibitem[Carrasco et~al.(2016)Carrasco, Turner, and Heath]{CARRASCO20161}
J.~Carrasco, M.~C. Turner, and W.~P. Heath.
\newblock {Zames–Falb} multipliers for absolute stability: From {O'Shea's}
  contribution to convex searches.
\newblock \emph{Euro. J. Control}, 28:\penalty0 1--19, 2016.

\bibitem[Cheng et~al.(2006)Cheng, Lin, and Shin]{multistableNNs}
C.-Y. Cheng, K.-H. Lin, and C.-W. Shin.
\newblock Multistability in recurrent neural networks.
\newblock \emph{SIAM J. Applied Math.}, 66\penalty0 (4):\penalty0 1301--1320,
  2006.

\bibitem[Cohen and Grossberg(1983)]{cohen_NN}
M.~A. Cohen and S.~Grossberg.
\newblock Absolute stability of global pattern formation and parallel memory
  storage by competitive neural networks.
\newblock \emph{IEEE Trans. Systems, Man, and Cybernetics}, SMC-13\penalty0
  (5):\penalty0 815--826, 1983.

\bibitem[Dalin et~al.(2022)Dalin, Ofir, Shalom, Ovseevich, Bullo, and
  Margaliot]{Dalin2022Duality_kcont}
O.~Dalin, R.~Ofir, E.~B. Shalom, A.~Ovseevich, F.~Bullo, and M.~Margaliot.
\newblock Verifying $k$-contraction without computing $k$-compounds, 2022.
\newblock URL \url{https://arxiv.org/abs/2209.01046}.
\newblock Submitted.

\bibitem[Davydov et~al.(2022)Davydov, Proskurnikov, and Bullo]{AD-AVP-FB:21k}
A.~Davydov, A.~V. Proskurnikov, and F.~Bullo.
\newblock {Non-Euclidean} contractivity of recurrent neural networks.
\newblock In \emph{American Control Conference}, Atlanta, USA, May 2022.

\bibitem[D\"{o}rfler and Bullo(2012)]{Dorfler2012kuramoto}
F.~D\"{o}rfler and F.~Bullo.
\newblock Synchronization and transient stability in power networks and
  nonuniform kuramoto oscillators.
\newblock \emph{SIAM J.\ Control Optim.}, 50\penalty0 (3):\penalty0 1616--1642,
  2012.

\bibitem[Fallat and Johnson(2011)]{total_book}
S.~M. Fallat and C.~R. Johnson.
\newblock \emph{Totally Nonnegative Matrices}.
\newblock Princeton University Press, Princeton, NJ, 2011.

\bibitem[Fiedler(2008)]{fiedler_book}
M.~Fiedler.
\newblock \emph{Special Matrices and Their Applications in Numerical
  Mathematics}.
\newblock Dover Publications, Mineola, NY, 2 edition, 2008.

\bibitem[Forni and Sepulchre(2019)]{Forni2019diff_diss}
F.~Forni and R.~Sepulchre.
\newblock Differential dissipativity theory for dominance analysis.
\newblock \emph{IEEE Transactions on Automatic Control}, 64\penalty0
  (6):\penalty0 2340--2351, 2019.

\bibitem[Gantmacher(1960)]{Gantmacher_vol1}
F.~R. Gantmacher.
\newblock \emph{The Theory of Matrices}, volume~I.
\newblock Chelsea Publishing Company, 1960.

\bibitem[Giaccagli et~al.(2022)Giaccagli, Andrieu, Tarbouriech, and
  Astolfi]{control_syn_vincent}
M.~Giaccagli, V.~Andrieu, S.~Tarbouriech, and D.~Astolfi.
\newblock Infinite gain margin, contraction and optimality: an {LMI}-based
  design.
\newblock \emph{Euro. J. Control}, page 100685, 2022.

\bibitem[Grussler and Sepulchre(2022)]{grussler2022variation}
C.~Grussler and R.~Sepulchre.
\newblock Variation diminishing linear time-invariant systems.
\newblock \emph{Automatica}, 136:\penalty0 109985, 2022.

\bibitem[Hopfield(1982)]{hopfield_net}
J.~J. Hopfield.
\newblock Neural networks and physical systems with emergent collective
  computational abilities.
\newblock \emph{Proceedings of the National Academy of Sciences}, 79\penalty0
  (8):\penalty0 2554–2558, 1982.

\bibitem[Horn and Johnson(1991)]{Horn1991TopicsMatrixAna}
R.~A. Horn and C.~R. Johnson.
\newblock \emph{Topics in Matrix Analysis}.
\newblock Cambridge University Press, 1991.

\bibitem[Khalil(2002)]{khalil_book}
H.~K. Khalil.
\newblock \emph{Nonlinear Systems}.
\newblock Prentice-Hall, Upper Saddle River, NJ, 3 edition, 2002.

\bibitem[Krotov and Hopfield(2016)]{krotov2016}
D.~Krotov and J.~J. Hopfield.
\newblock Dense associative memory for pattern recognition.
\newblock In D.~Lee, M.~Sugiyama, U.~Luxburg, I.~Guyon, and R.~Garnett,
  editors, \emph{Advances in Neural Information Processing Systems}, volume~29.
  Curran Associates, Inc., 2016.

\bibitem[Li and Muldowney(1995)]{li1995}
M.~Y. Li and J.~S. Muldowney.
\newblock On {R. A. Smith's} autonomous convergence theorem.
\newblock \emph{Rocky Mountain J. Math.}, 25\penalty0 (1):\penalty0 365--378,
  1995.

\bibitem[Li et~al.(1999)Li, Graef, Wang, and Karsai]{LI1999191}
M.~Y. Li, J.~R. Graef, L.~Wang, and J.~Karsai.
\newblock Global dynamics of a {SEIR} model with varying total population size.
\newblock \emph{Mathematical Biosciences}, 160\penalty0 (2):\penalty0 191--213,
  1999.

\bibitem[Lohmiller and Slotine(1998)]{LOHMILLER1998683}
W.~Lohmiller and J.-J.~E. Slotine.
\newblock On contraction analysis for non-linear systems.
\newblock \emph{Automatica}, 34:\penalty0 683--696, 1998.

\bibitem[Margaliot(2006)]{MARGALIOT20062059}
M.~Margaliot.
\newblock Stability analysis of switched systems using variational principles:
  An introduction.
\newblock \emph{Automatica}, 42\penalty0 (12):\penalty0 2059--2077, 2006.

\bibitem[Margaliot and Sontag(2019)]{margaliot2019revisiting}
M.~Margaliot and E.~D. Sontag.
\newblock Revisiting totally positive differential systems: A tutorial and new
  results.
\newblock \emph{Automatica}, 101:\penalty0 1--14, 2019.

\bibitem[Megretski and Rantzer(1997)]{ICQ}
A.~Megretski and A.~Rantzer.
\newblock System analysis via integral quadratic constraints.
\newblock \emph{IEEE Trans.\ Automat.\ Control}, 42\penalty0 (6):\penalty0
  819--830, 1997.

\bibitem[Miranda-Villatoro et~al.(2018)Miranda-Villatoro, Forni, and
  Sepulchre]{MIRANDAVILLATORO201876}
F.~A. Miranda-Villatoro, F.~Forni, and R.~J. Sepulchre.
\newblock Analysis of {Lur’e} dominant systems in the frequency domain.
\newblock \emph{Automatica}, 98:\penalty0 76--85, 2018.

\bibitem[Muldowney(1990)]{muldowney1990compound}
J.~S. Muldowney.
\newblock Compound matrices and ordinary differential equations.
\newblock \emph{Rocky Mountain J. Math.}, 20\penalty0 (4):\penalty0 857--872,
  1990.

\bibitem[Ofir and Margaliot(2021)]{ron_DAE}
R.~Ofir and M.~Margaliot.
\newblock The multiplicative compound of a matrix pencil with applications to
  difference-algebraic equations.
\newblock 2021.
\newblock URL \url{arXiv preprint arXiv:2111.01419}.
\newblock Submitted.

\bibitem[Ofir et~al.(2022{\natexlab{a}})Ofir, Margaliot, Levron, and
  Slotine]{ofir2021sufficient}
R.~Ofir, M.~Margaliot, Y.~Levron, and J.-J. Slotine.
\newblock A sufficient condition for $k$-contraction of the series connection
  of two systems.
\newblock \emph{IEEE Trans.\ Automat.\ Control}, 67\penalty0 (9):\penalty0
  4994--5001, 2022{\natexlab{a}}.

\bibitem[Ofir et~al.(2022{\natexlab{b}})Ofir, Ovseevich, and
  Margaliot]{ron_ifac_version}
R.~Ofir, A.~Ovseevich, and M.~Margaliot.
\newblock A sufficient condition for k-contraction in {Lurie} systems.
\newblock In \emph{IFAC 2023 World Congress}, 2022{\natexlab{b}}.
\newblock Submitted.

\bibitem[Proskurnikov et~al.(2022)Proskurnikov, Davydov, and
  Bullo]{Proskurnikov2022GeneralizedSLemma}
A.~V. Proskurnikov, A.~Davydov, and F.~Bullo.
\newblock The {Yakubovich} {S-Lemma} revisited: Stability and contractivity in
  non-{Euclidean} norms, 2022.
\newblock URL \url{https://arxiv.org/abs/2207.14579}.

\bibitem[Qiao et~al.(2001)Qiao, Peng, and Xu]{cont_hopfield}
H.~Qiao, J.~Peng, and Z.-B. Xu.
\newblock Nonlinear measures: a new approach to exponential stability analysis
  for {Hopfield}-type neural networks.
\newblock \emph{IEEE Trans. Neural Networks}, 12\penalty0 (2):\penalty0
  360--370, 2001.

\bibitem[Sauer and Pai(1998)]{Sauer1998power}
P.~Sauer and M.~Pai.
\newblock \emph{Power System Dynamics and Stability}.
\newblock Prentice Hall, 1998.
\newblock ISBN 9780136788300.

\bibitem[Schwarz(1970)]{schwarz1970}
B.~Schwarz.
\newblock Totally positive differential systems.
\newblock \emph{Pacific J. Math.}, 32\penalty0 (1):\penalty0 203--229, 1970.

\bibitem[Smith(1986)]{Smith1986haus}
R.~A. Smith.
\newblock Some applications of {Hausdorff} dimension inequalities for ordinary
  differential equations.
\newblock \emph{Proc. Royal Society of Edinburgh: Section A Mathematics},
  104\penalty0 (3-4):\penalty0 235--259, 1986.

\bibitem[Str{\"o}m(1975)]{strom1975logarithmic}
T.~Str{\"o}m.
\newblock On logarithmic norms.
\newblock \emph{SIAM J. Numerical Analysis}, 12\penalty0 (5):\penalty0
  741--753, 1975.

\bibitem[Vidyasagar(2002)]{vidyasagar2002nonlinear}
M.~Vidyasagar.
\newblock \emph{Nonlinear Systems Analysis}.
\newblock SIAM, 2002.

\bibitem[Weiss et~al.(2019)Weiss, D\"{o}rfler, and
  Levron]{Weiss2019StabilityNRPS}
G.~Weiss, F.~D\"{o}rfler, and Y.~Levron.
\newblock A stability theorem for networks containing synchronous generators.
\newblock \emph{Systems \& Control Letters}, 134:\penalty0 104561, 2019.

\bibitem[Wu and Margaliot(2022)]{wu2021diagonal}
C.~Wu and M.~Margaliot.
\newblock Diagonal stability of discrete-time k-positive linear systems with
  applications to nonlinear systems.
\newblock \emph{IEEE Trans.\ Automat.\ Control}, 67\penalty0 (8):\penalty0
  4308--4313, 2022.

\bibitem[Wu et~al.(2022{\natexlab{a}})Wu, Kanevskiy, and Margaliot]{kordercont}
C.~Wu, I.~Kanevskiy, and M.~Margaliot.
\newblock $k$-contraction: theory and applications.
\newblock \emph{Automatica}, 136:\penalty0 110048, 2022{\natexlab{a}}.

\bibitem[Wu et~al.(2022{\natexlab{b}})Wu, Pines, Margaliot, and
  Slotine]{wu2020generalization}
C.~Wu, R.~Pines, M.~Margaliot, and J.-J. Slotine.
\newblock Generalization of the multiplicative and additive compounds of square
  matrices and contraction theory in the {Hausdorff} dimension.
\newblock \emph{IEEE Trans.\ Automat.\ Control}, 67\penalty0 (9):\penalty0
  4629--4644, 2022{\natexlab{b}}.

\end{thebibliography}

 \noindent \textbf{Ron Ofir} %(Student Member, IEEE) 
    received his BSc degree (cum laude) in Elec.  Eng. from the Technion-Israel Institute of Technology in~2019. He is currently pursuing his Ph.D. degree in the  Andrew and Erna Viterbi Faculty of Electrical and Computer Engineering, Technion - Israel Institute of Technology. His current research interests include compound matrices and
 contraction theory with
 applications in  dynamics and control of power systems.

\subsection*{  } % This subsection (with no heading) is added to give more space between two biographies
    \setlength\intextsep{0pt} % align top of photo with text
    \begin{wrapfigure}{l}{0.14\textwidth}
        \centering
        \includegraphics[width=0.15\textwidth]{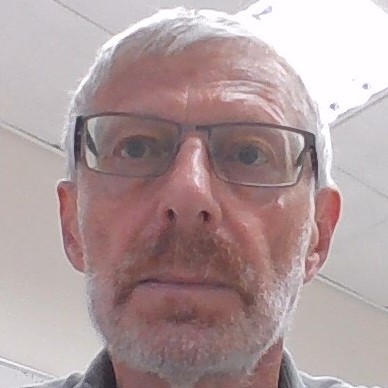} 
    \end{wrapfigure}
    \noindent \textbf{Alexander Ovseevich}
    received his MSc in Mathematics from the Moscow State University in 1973, and PhD in Mathematics from the Leningrad division of the Steklov mathematical institute, RAS in 1976. He got a degree of the doctor of Phys.-Math. Sciences from the Institute for Problems in Mechanics, Moscow, Russia in 1997. He worked in the Institute for Problems in Mechanics from 1978 to 2022 as a senior research fellow and a leading research fellow. In 2022 he joined the Dept. of Elec. Eng.-Systems, Tel Aviv University, where he is currently a research assistant. His research interests include many topics from Number Theory to Mathematical Physics and Control Theory.

\subsection*{  } % This subsection (with no heading) is added to give more space between two biographies
    \setlength\intextsep{0pt} % align top of photo with text
    \begin{wrapfigure}{l}{0.13\textwidth}
        \centering
        \includegraphics[width=0.15\textwidth]{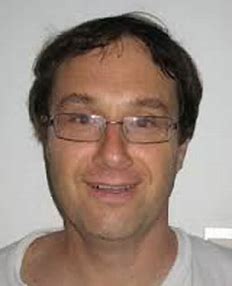}
    \end{wrapfigure}
    \noindent \textbf{Michael Margaliot}  received the BSc (cum laude) and MSc degrees in
 Elec. Eng. from the Technion-Israel Institute of Technology-in
 1992 and 1995, respectively, and the PhD degree (summa cum laude) from Tel
 Aviv University in 1999. He was a post-doctoral fellow in the Dept. of
 Theoretical Math. at the Weizmann Institute of Science. In 2000, he
 joined the Dept. of Elec. Eng.-Systems, Tel Aviv University,
 where he is currently a Professor. His  research
 interests include the stability analysis of differential inclusions and
 switched systems, optimal control theory, computation with
 words, Boolean control networks, contraction theory,  applications of matrix compounds in systems and control theory, and systems biology.
 He is co-author of \emph{New Approaches to Fuzzy Modeling and Control: Design and
 Analysis}, World Scientific,~2000 and of \emph{Knowledge-Based Neurocomputing}, Springer,~2009. 
 He  served
as  an Associate Editor of~\emph{IEEE Trans. on Automatic Control} during 2015-2017.

\end{document}